\documentclass[twocolumn,numberedappendix]{aastex63}
\usepackage{apjfonts}
\usepackage{epsfig}
\usepackage{amsmath}

\newcommand{\G}{\Gamma}

\newcommand{\sT}{\sigma_{\rm T}}
\newcommand{\p}{^\prime}
\newcommand{\e}{\epsilon}
\newcommand{\g}{\gamma}
\newcommand{\gp}{\gamma^{\prime}}

\newcommand{\ep}{\epsilon^\prime}

\newcommand{\dD}{\delta_{\rm D}}

\newcommand{\psim}{\lower.5ex\hbox{$\; \buildrel \propto \over\sim \;$}}
\newcommand{\lbar}{\lower.0ex\hbox{$\; \buildrel
{\lower0.0ex \hbox{-}} \over\lambda  \;$}}

\newcommand{\cm}{\mathrm{cm}}

\newcommand{\erg}{\mathrm{erg}}

\newcommand{\s}{\mathrm{s}}

\newcommand{\Ngreen}{N_e}

\shorttitle{Electron Acceleration in Blazars}
\shortauthors{Lewis, Finke, \& Becker}

\begin{document}
\title{Electron Acceleration in Blazars: Application to the 3C~279 Flare on 2013 December 20}

\author[0000-0002-9854-1432]{Tiffany R.\ Lewis}
\affil{Haifa Research Center for Theoretical Physics \& Astrophysics, University of Haifa, Mt. Carmel. Haifa 3498838, Israel}
\affil{	Department of Astrophysics, Faculty of Exact Sciences, Tel Aviv University, P.O. Box 39040, Tel Aviv 6997801, Israel}
\email{tiffanylewisphd@gmail.com}

\author{Justin D.\ Finke}
\affil{U.S.\ Naval Research Laboratory, Code 7653, 4555 Overlook Ave.\ SW,
        Washington, DC,
        20375-5352}

\author{Peter A.\ Becker}
\affil{	Department of Physics \& Astronomy, MSN 3F3,
	George Mason University, 4400 University Drive, Fairfax, VA 22030}

\submitjournal{The Astrophysical Journal}
\accepted{September 9, 2019}

\begin{abstract}
The broadband spectrum from the 2013 December 20 $\g$-ray flare from
3C~279 is analyzed with our previously-developed one-zone blazar jet
model.  We are able to reproduce two SEDs, a quiescent and flaring state, the latter of which
had an unusual SED, with hard $\g$-ray spectrum, high Compton
dominance, and short duration.  Our model suggests that there is insufficient energy for a
comparable X-ray flare to have occurred simultaneously, which is an
important constraint given the lack of X-ray data. We show that
first- and second-order Fermi acceleration are sufficient to explain the
flare, and that magnetic reconnection is not needed. The model includes 
particle acceleration, escape, and adiabatic and radiative energy losses, including the full Compton
cross-section, and emission from the synchrotron, synchrotron
self-Compton, and external Compton processes.   We provide a
simple analytic approximation to the electron distribution solution to
the transport equation that may be useful for simplified modeling in
the future.
\end{abstract}


\section{Introduction}
\label{intro}


Blazars, the most energetic sustained phenomena in the known Universe, are radio-loud active galactic nuclei with relativistic jets closely aligned with our line of sight. The broadband spectral energy distributions (SEDs) from blazars are dominated by beamed jet emission. Blazars are known for their variable broadband spectral features, extending from radio through $\gamma$-rays. Their broadband SEDs are characterized by two primary features.  In leptonic models, those include a synchrotron peak at lower energies and a Compton peak at higher energies.  Alternatively, the high-energy peak can be produced by proton synchrotron \citep[e.g.][]{aharonian00, mucke03, reimer04} or the decay products of proton-photon interactions \citep[e.g.][]{sikora87,mannheim92,protheroe95}, however hadronic processes are disfavored in some cases due to the excessive energy requirements \citep[e.g.][]{boett13, zdziarski15,petropoulou16}. Blazars are divided into flat-spectrum radio quasars (FSRQs) and BL Lac objects based on their optical spectra, with the former having strong broad emission lines, while the latter do not. FSRQs are thought to have their $\g$-ray emission dominated by the external Compton process, where the seed photon fields for Compton scattering come from outside the jet, from sources such as the broad-line region \citep[BLR; e.g.,][]{sikora94,blandford95,ghisellini96}, dust torus \citep[e.g.,][]{kataoka99,blazejowski00}, and accretion disk \citep[e.g.,][]{dermer92,dermer93}.

Blazars are characterized by stochastic variability \citep[e.g.,][hereafter Paper~1]{finke14,finke15,lewis16}. For example, the FSRQ 3C 279 alternates between quiescence and flaring states over periods of a few days to several weeks \citep[e.g.][]{hayash12,hayash15}.  Additionally, FSRQs can exhibit isolated flaring activity at optical, X-ray, or $\g$-ray energies \citep[e.g.][]{ostermeyer09, hayash12,macdonald17}, or correlated flares \citep[e.g.][]{hayash12,marscher12,liodakis18}.  A single blazar can also exhibit different types of flares at different times \citep[e.g.][and references therein]{hayash12,vittorini14,kaur18}, and currently there is no consensus on an explanation for why some flares exhibit correlated variability and some do not.

Particle acceleration is a necessity in astrophysical jets in order to explain how the radiating particles reach the energies required to produce the observed emission. However, currently there is no consensus on the precise combination of required acceleration mechanisms, or on their effects on the observed blazar spectra \citep[e.g.][]{madejski16,romero17}.  The first-order Fermi process, resulting from acceleration at a shock front, is widely assumed to contribute to particle acceleration in blazars \citep[e.g.][]{bednarek97,tavecchio98,bednarek99,finke08_SSC,dermer09,hayash12}. On its own,  first-order Fermi acceleration produces a power-law particle spectrum \citep{fermi49}. This can be attenuated by radiative loss mechanisms, especially synchrotron, which impose an exponential cutoff at high energies. Second-order Fermi (stochastic) acceleration is also thought to contribute to electron energization in blazars \citep[e.g.][]{summerlin12,baring17}.  Stochastic acceleration alone gives a log-parabola electron distribution (ED) \citep[e.g.][]{tramacere11}. 

We previously developed a steady-state model for particle acceleration and emission in blazar jets which included first- and second-order Fermi  acceleration of particles, and radiation by synchrotron, synchrotron self-Compton (SSC), and external Compton (EC) mechanisms \citep[][hereafter Paper~2]{lewis18}.  EC of photons from a realistic, stratified BLR \citep{finke16} and dust torus were included.  Here we exercise this model in a new situation: modeling the extreme flare from 3C 279 observed on 2013 December 20.

The familiar blazar 3C 279 has been very well observed at all
wavelengths for many years
\citep[e.g.][]{grandi96,wehrle98,chatterjee08,hayash12}.  On 2013 December
20, the source was observed by the {\em Fermi} Large Area Telescope
(LAT) to have a flare that was extremely hard in $\g$ rays
(photon spectral index $\G_{\g}\sim1.7$, compared to 2.4 right before the flare), with the
peak of the $\g$-ray emission shifting from below the {\em Fermi}-LAT
energy range ($h\nu^{\rm peak}\lesssim 100$\ MeV) immediately before
the flare, to over a decade in energy higher, $h\nu^{\rm
  peak}\gtrsim2$\ GeV \citep{hayash15}.  At the same time, the peak of
the synchrotron spectrum stayed nearly constant, and may have even
shifted to slightly higher frequencies; the Compton dominance was
estimated to be an extreme $A_{\rm C} = L_{\g}/L_{\rm syn}\gtrsim300$.
The flux doubling timescale was rapid, $\approx 2\,$hr, although still
less extreme than the minute-scale variability found by
\citet{ackermann16} for the 2015 June flare from 3C 279.

The 2013 December 20 flare has attracted some attention from modelers due
to its unusual properties \citep[e.g.,][]{hayash15,asano15,paliya16},
although no firm conclusions regarding the particle acceleration
mechanism have been drawn.  Here we apply our steady-state blazar
acceleration and emission model to this problem.

The paper is organized as follows. In Section~\ref{model}, we give a brief
summary of our jet model, which includes a steady-state, self-consistent electron transport equation
(including first- and second order Fermi acceleration; synchrotron and Compton
losses). In Section~\ref{3c279},
we apply the model to the extreme 2013 December 20 $\g$-ray flare of 3C~279
(as well as the preceding quiescent period) to obtain new physical
insights. We discuss and interpret the results of the analysis in
Section \ref{discuss}.  We include a derivation of the simplified
analytic electron distribution (ED) in Appendix \ref{ap-bump}.  In Appendix \ref{ap-deps}, we
provide derivations related to the physical interpretation of the ED
shape.

\section{Model}
\label{model}


The blazar jet originates from a supermassive black hole (BH), perhaps
accelerated by a rapidly spinning BH threaded with magnetic fields
anchored in an accretion disk \citep{blandford77}.  The jet plasma
moves outward from the BH, and toward the observer, with some bulk
Lorentz factor $\G = (1-\beta^2)^{-1/2}$, which is related to the
relativistic bulk speed $v=\beta c$, where $c$ is the speed of light.
The material moves relativistically toward the observer, within some
small angle $\theta$ to the line of sight, leading to a Doppler factor
$\dD=[\G(1-\beta\cos\theta)]^{-1}$. We assume $\dD=\G$.

The primary emitting region is modeled as a single, compact
homogeneous zone or ``blob.'' The co-moving blob is causally connected
by the light crossing timescale $t_{\rm var}$, which is the minimum
variability timescale in the observer's frame.  Thus, the radius of the blob (in the frame
co-moving with the jet) must be $R\p_b \lesssim c\,\dD t_{\rm
  var}/(1+z)$, where $z$ denotes the cosmological redshift of the
source.  Radio emission is produced throughout the jet via synchrotron
emission \citep[e.g.][]{blandford79,konigl81,finke19}.  The relatively small size of the blob radius implied by the observed variability timescales suggests that significant synchrotron self-absorption occurs in the blob, making it unlikely that the blob is the source of the observed radio emission.  The radio emission thus provides upper limits on the emission from the region considered here.

\subsection{Electron Distribution}
\label{EDsection}

Throughout the data comparison process, we use the numerical model, including the self-consistent ED (Paper~2). 
We describe the ED $N_e(\g)$ in the frame of the blob using
a steady-state Fokker-Planck equation,
\begin{align}
0 = & \frac{\partial^2}{\partial \gamma^2} \left( \frac{1}{2}
\frac{d \sigma^2}{d t} \Ngreen \right) - \frac{\partial}{\partial \gamma}
\left( \left< \frac{d \gamma}{d t} \right> \Ngreen \right) \nonumber \\ 
& - \frac{\Ngreen}{t_{\rm esc}} + \dot{N}_{\rm inj} \delta(\gamma-\gamma_{\rm inj})  \ ,
\label{eq-transport}
\end{align}
where $\g \equiv E/(m_ec^2)$ is the electron Lorentz factor, $m_e$ is
the electron mass, and $c$ is the speed of light in a vacuum.  Since
this is a steady state equation, we have set $\partial
\Ngreen/\partial t = 0$.  The acceleration and synchrotron energy loss timescales are appreciably shorter than the observed rise time of the flare (see Section \ref{timescales}), making the steady-state calculation appropriate in this case.  We note that we use the term ``electrons''
here and throughout this paper to refer to both electrons and
positrons.

In Equation (\ref{eq-transport}), the
energy-dependent particle escape timescale $t_{\rm esc}$, is related to the dimensionless escape parameter $\tau$ via
\begin{equation}
t_{\rm esc}(\g) = \frac{\tau}{D_0 \gamma} \ ,
\end{equation}
where
\begin{equation}
\tau \equiv \frac{R^{\prime 2}_{b}qBD_0}{m_ec^3} \ ,
\end{equation}
in the Bohm limit (Paper~1), and $q$ is the fundamental charge.  The Lorentz factor of the injected electrons is $\g_{\rm inj}$ in the particle injection rate
\begin{equation}
\dot{N}_{\rm inj} = \frac{L_{\rm e,inj}}{m_ec^2\g_{\rm inj}}\ ,
\end{equation}
where $L_{\rm e,inj}$ is the electron injection luminosity.  Both
$\g_{\rm inj}$ and $L_{\rm e,inj}$ are implemented as free parameters, but the former is held constant in the subsequent analysis at the lower numerical grid limit to simulate a thermal particle source.  We solved
equation (\ref{eq-transport}) using the full Compton energy loss rate
numerically, as described in Paper~2.

In Equation (\ref{eq-transport}), the broadening coefficient 
\begin{equation}
\frac{1}{2}\frac{d \sigma^2}{d t} = D_0 \, \g^2 \ ,
\label{eq-broadcoef}
\end{equation}
is consistent with hard-sphere scattering, where
$D_0\propto \s^{-1}$ \citep{park95} is a free parameter.  The drift coefficient expresses the mean rate at which each process contributes 
\begin{equation}
\left< \frac{d \gamma}{d t} \right> = 
D_0 \left[4\gamma+a\gamma-b_{\rm syn}\gamma^2 - 
\gamma^2 \sum_{j=1}^J b^{(j)}_{\rm C} H(\gamma \epsilon_{\rm ph}^{(j)}) \right] \ ,
\label{eq-driftKN}
\end{equation}
where second-order Fermi acceleration occurs at a rate
\begin{equation}
\dot{\g}_{\rm sto} = 4 D_0 \g \ ,
\label{eq-StoRate}
\end{equation}
and 
\begin{equation}
\dot{\g}_{\rm ad+sh} \equiv a D_0 \g \equiv A_{\rm ad+sh} \g \ ,
\label{eq-AdShRate}
\end{equation}
where $a$ is a dimensionless free parameter that includes first-order Fermi
acceleration and adiabatic cooling\footnote{For $a>0$ first-order Fermi
  acceleration dominates over adiabatic losses; for $a<0$, the
  opposite is true.}.
The coefficients $A_{\rm sh}$ and $A_{\rm ad}$ represent the first-order Fermi acceleration rate and the adiabatic loss rate, respectively.
The rate of synchrotron cooling is
\begin{equation}
|\dot{\g}_{\rm syn}| \equiv D_0 b_{\rm syn} \gamma^2 =
\frac{\sigma_{\rm T}B^2 }{6 \pi m_e c} \g^2 \ ,
\label{eq-SynRate}
\end{equation}
where $\sT=6.65\times10^{-25}$ cm$^2$ is the Thomson cross-section and
$B$ is the strength of the tangled, homogeneous magnetic field. Similarly,
the Compton cooling rate for each individual component is
\begin{align}
\label{eq-CompRate}
|\dot{\gamma}_{\rm C}|  \equiv & D_0\gamma^2  b_{\rm C}^{(j)} 
H(\gamma \epsilon_{\rm ph}^{(j)}) 
\equiv \g^2   B_{\rm C}^{(j)} H(\gamma \epsilon_{\rm ph}^{(j)})  \\ &=
\g^2   \frac{4 \sigma_{\rm T} \G^2u_{\rm ph}^{(j)}}{3 m_e c }
H(\gamma \epsilon_{\rm ph}^{(j)})   \ . \nonumber
\end{align}
Here $b_{\rm C}^{(j)}$ is a dimensionless constant related to
Compton cooling for the different external radiation fields $j$, with
energy densities $u_{\rm ph}^{(j)}$. The function $H(y)$ is a
complicated expression related to mitigation of energy losses with the
full Compton cross-section \citep{boett97}.  We include a dust torus
and 26 broad lines, for a total of $J=27$ EC components
\citep{finke16}.

\subsection{Thomson regime approximation}
\label{EDappx}

In the Thomson limit,
$y\ll 1$, $H(y)\rightarrow 1$, the steady state Fokker-Planck equation
(Equation [\ref{eq-transport}]) has the analytic solution (see Paper~1
for details)
\begin{align}
\label{eq-analytic}
N_e(\g) & = \frac{\dot{N}_{\rm inj}}{bD_0} \frac{\G(\mu - \kappa + 1/2)}{\G(1+2\mu)} e^{(\g_{\rm inj}-\g) b/2} \g_{\rm inj}^{-2-a/2} \g^{a/2} \nonumber \\ & \times
\begin{cases}
{\mathcal M}_{\kappa,\mu}(b \g) {\mathcal W}_{\kappa, \mu} (b\g_{\rm inj}) & , \g \le \g_{\rm inj} \\
{\mathcal M}_{\kappa,\mu}(b \g_{\rm inj}) {\mathcal W}_{\kappa, \mu} (b\g) & , \g_{\rm inj} \le \g
\end{cases} \ , 
\end{align}
where $b \equiv b_{\rm syn}+\sum_{j=1}^J  b_{\rm C}^{(j)}$, ${\mathcal M}_{\kappa,\mu}$ and ${\mathcal W}_{\kappa,\mu}$ 
are Whittaker functions \citep[][]{slater60}, with coefficients
\begin{equation}
\kappa=2 - \frac{1}{b\tau} + \frac{a}{2},\quad 
{\rm and } 
\quad 
\mu=\frac{a+3}{2} \ .
\label{eq-lambda-sigma}
\end{equation}

We found the analytic Thomson regime solution, Equation (\ref{eq-analytic}), 
previously (Paper~1).  Here we provide a useful approximation, 
\begin{align}
\label{eq-anasim}
N^{\rm app}_e(\g) & \approx \frac{\dot{N}_{\rm inj}}{D_0} e^{-b\g} 
\bigg[  \frac{\tau b^{a+4} \g^{a+2}}{\G(a+4)} +\frac{1}{3+a}\frac{1}{\g} \bigg]
\end{align}
in the $\g_{\rm inj}<\g$ regime (where most of our analysis takes place), and where
$b\g \ll 1$ and $(b\tau)^{-1} \ll 1$ are the
simplifying assumptions, which are valid in all blazar analyses we have examined, both here and in Paper~2. The precision of these assumptions is addressed further in Appendix \ref{ap-bump}, however the simplified solution is relatively accurate for most applications, except where $(b\tau)^{-1} \gtrsim 1$.

The constraint $b\g \ll 1$ indicates that the rate of particle energy lost to synchrotron (Equation \ref{eq-SynRate}) and Thomson processes (Equation \ref{eq-CompRate}, for $H(\epsilon\g)=1$), where ($b = b_{\rm syn} + b^{(j)}_{\rm C}$) must be smaller than the rate at which energy is gained by the second-order Fermi process (Equation \ref{eq-StoRate}). Thus, the emitting blob must be in the acceleration region, which can be demonstrated by the energy budget for the flare presented in this work (Section \ref{budget}).  Similarly, the requirement $(b\tau)^{-1} \ll 1$ is analogous to the expectation that the energy losses due to the synchrotron  and Thomson processes outpace energy lost due to particle escape (i.e. the fast-cooling regime), which is apparent in the present application by examining the energy budget (Section \ref{budget}).

\begin{figure}[t]
\vspace{2.2mm} 
\centering
\includegraphics[width=0.45\textwidth]{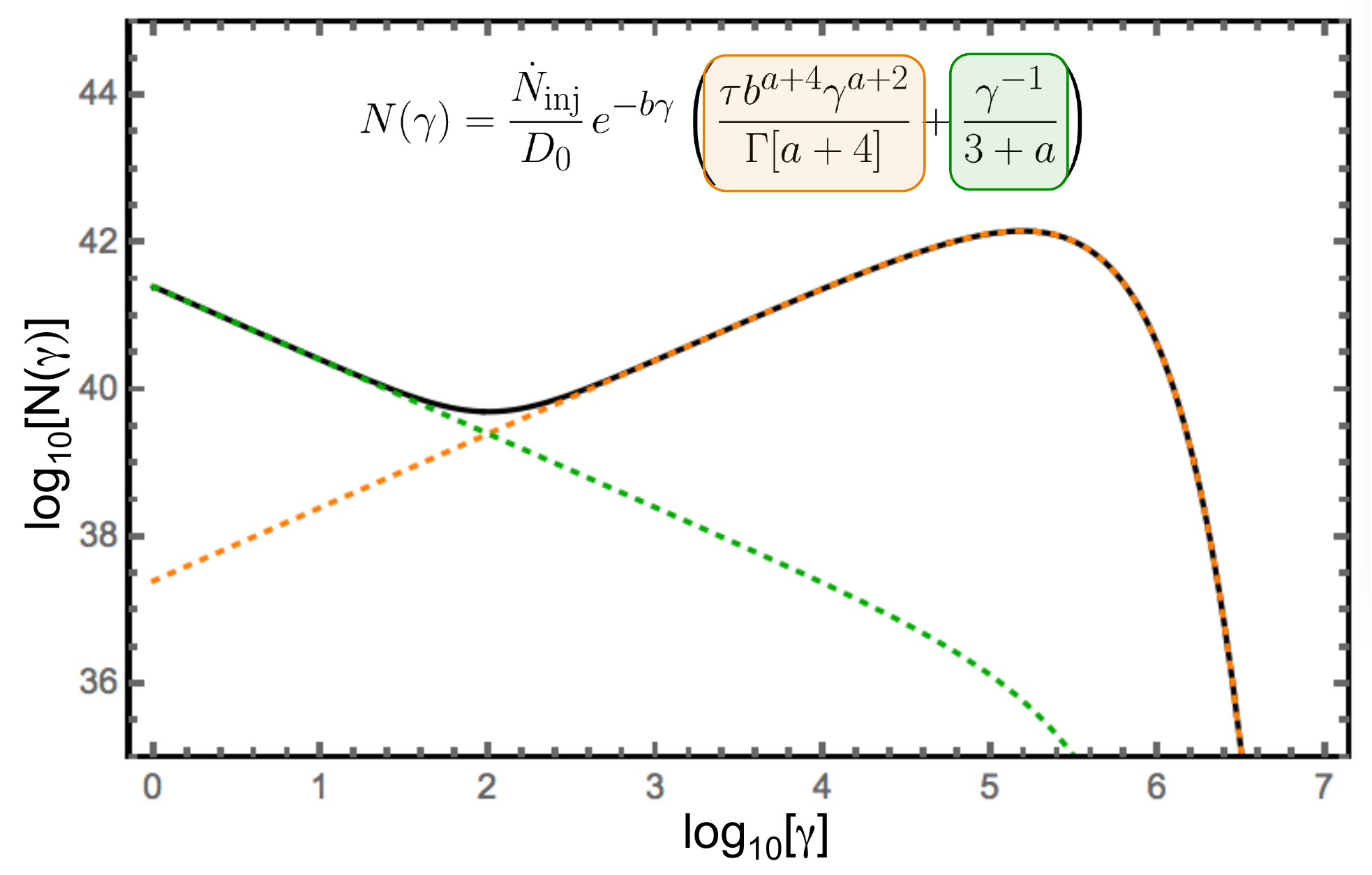} 
\caption{This sample ED uses contrived parameters to showcase the two components from the $\g \ge \g_{\rm inj}$ branch of Equation (\ref{eq-anasim}).  The orange line is given by a balance between the the first-order Fermi/adiabatic and second-order acceleration terms (Appendix \ref{ap-deps}), and $a>-2$, giving a positive power law, and an excess of high-energy particles.  The green line is dominated by second-order Fermi (stochastic) acceleration of particles to higher energies (Appendix \ref{ap-deps}). Both terms are independently affected by the exponential cutoff due to emission mechanisms. The black line is the full Thomson regime solution (Equation \ref{eq-analytic}).} 
\label{fig-ElecDist_Simp}
\vspace{2.2mm}
\end{figure}

The full derivation and complementing solution for $\g>\g_{\rm inj}$ are given in Appendix
\ref{ap-bump}.  The main shape of the ED is governed by two
components: one is driven by a balance between first- and second-order Fermi acceleration($\propto \g^{a+2}$;
orange curve in Figure \ref{fig-ElecDist_Simp}) and the other is due to second-order Fermi acceleration ($\propto \g^{-1}$;
green curve in Figure \ref{fig-ElecDist_Simp}; see Appendix
\ref{ap-deps}).  The ratio of first- to second-order Fermi
acceleration acceleration (parameterized by $a = A_{\rm ad+sh}/D_0$) impacts
the shared term. When $a < -2$, the lower-energy power law is
negative, and when $a > -2$, the $\propto \g^{a+2}$ term is increasing
with increasing energy.  In practice, the shape of the the term does
not vary significantly for most analyses, but is important in simulating the flare examined here.  The exponential cutoff is governed by the
emission mechanisms, which are constrained to the Thomson limit for
the analytic solution.  Appendix \ref{ap-bump} discusses the interpretation of the acceleration mechanisms in the
simplified solution in further detail.

The analytic solution and others derived from it, are useful for
physical interpretation of the shape of the ED with
regard to acceleration mechanisms.  However, all of the data interpretation
here is performed with the full numerical model using the full Compton
cross-section and energy losses.

\subsection{Emission}
\label{emission}

We include thermal emission from the accretion disk \citep{shakura73}
and dust torus, in addition to emission from the jet from synchrotron,
SSC, and EC of dust torus and BLR photons.  The source 3C 279 has
redshift $z=0.536$ giving it a luminosity distance
$d_L=9.6\times10^{27}\ \cm$ in a cosmology where $(h, \Omega_m,
\Omega_\Lambda) = (0.7, 0.3, 0.7)$.

The $\nu F_\nu$ disk flux is approximated as
\begin{flalign}
f^{\rm disk}_{\e_{\rm obs}} = \frac{1.12}{4\pi d_L^2}\ 
\left( \frac{ \e}{ \e_{\rm max}} \right)^{4/3} {\rm e}^{-\e / \e_{\rm max}}
\end{flalign}
\citep{dermer14} where $\e=\e_{\rm obs}(1+z)$ and $m_e c^2 \e_{\rm max} = 10$ eV.
The $\nu F_\nu$ dust torus flux is approximated as a blackbody, 
\begin{flalign}
f^{\rm dust}_{\e_{\rm obs}} = 
\frac{15 L^{\rm dust}}{4\pi^5d_L^2} \frac{(\e/\Theta)^4}{\exp(\e/\Theta) - 1} \ ,
\end{flalign}
where again $\e=\e_{\rm obs}(1+z)$, and also $\Theta = k_{\rm B}T_{\rm dust}/(m_ec^2)$, 
$T_{\rm dust}$ is the dust temperature, and $k_B$ is the Boltzmann constant.

The jet blob $\nu F_\nu$ flux is computed using the ED solution to
the electron Fokker-Planck equation (Equation [\ref{eq-transport}]), 
$N\p_e(\gp)$.  We now add primes to indicate the distribution is in the
frame co-moving with the jet blob.

The synchrotron flux
\begin{flalign}
f_{\epsilon_{\rm obs}}^{\rm syn} = 
\frac{\sqrt{3} \e' \delta_{\rm D}^4 e^3 B}{4\pi h d_{\rm L}^2} 
\int^\infty_1 d\gp\ N\p_e(\gp)\ R(x)\ ,
\label{fsy}
\end{flalign}
where 
\begin{flalign}
x = \frac{4\pi \epsilon' m_e^2 c^3}{3eBh\g^{\prime 2}}\ ,
\label{eq-x}
\end{flalign}
and $R(x)$ is defined by \citet{crusius86}.  Synchrotron
self-absorption is also included.  The SSC flux
\begin{flalign}
f_{\e_s}^{\rm SSC} & = \frac{9}{16} \frac{(1+z)^2 \sT \e_s^{\prime 2}}
{\pi \dD^2 c^2t_{v}^2 } 
 \int^\infty_0\ d\ep_*\ 
\frac{f_{\e_*}^{\rm syn}}{\e_*^{\prime 3}}\ 
\nonumber \\ & \times
\int^{\infty}_{\gp_{1}}\ d\gp\ 
\frac{N\p_e(\gp)}{\g^{\prime2}} F_{\rm C}\left(4 \g\p \ep_{*}, \frac{\e}{\g\p} \right) \ ,
\label{fSSC}
\end{flalign}
\citep[e.g.,][]{finke08_SSC} where $\ep_s= \e_s(1+z)/\dD$, $\ep_*=\e_*(1+z)/\dD$, and
\begin{flalign}
\gp_{1} = \frac{1}{2}\epsilon\p_s 
\left( 1+ \sqrt{1+ \frac{1}{\epsilon\p \epsilon\p_s}} \right) \ .
\end{flalign}
The function $F_{\rm C}(p,q)$ was originally derived by
\citet{jones68}, but had a mistake that was corrected by
\citet{blumen70}.  The EC flux \citep[e.g.,][]{georgan01,dermer09}
\begin{flalign}
\label{eq-ECflux}
f_{\e_s}^{\rm EC} & = \frac{3}{4} \frac{c\sT \e_s^2}{4\pi d_L^2}\frac{u_*}{\e_*^2} \dD^3 
\nonumber \\ & \times
\int_{\g_{1}}^{\g_{\max}} d\g
\frac{N\p_e(\g/\dD)}{\g^2}F_{\rm C} \left(4 \g \e_*, \frac{\e_s}{\g} \right) \ ,
\end{flalign}
where
\begin{flalign}
\g_{1} = \frac{1}{2}\e_s \left( 1+ \sqrt{1+ \frac{1}{\e \e_s}} \right) \ 
\end{flalign}
and $u_*$ and $\e_*$ are the 
energy density and dimensionless photon energy of the external radiation
field, respectively.  For the dust torus photons, 
\begin{equation}
u_* = u_{\rm dust} = 2.2 \times 10^{-5}  \bigg(\frac{ \xi_{\rm dust}}{0.1} \bigg) 
\bigg( \frac{T_{\rm dust}}{1000\ {\rm K}} \bigg)^{5.2}  ~ {\rm erg ~ cm^{-3}} \ 
\end{equation}
and
\begin{equation}
\e_* = \e_{\rm dust} = 5 \times 10^{-7} 
\bigg( \frac{T_{\rm dust}}{1000\ {\rm K}} \bigg) \ ,
\end{equation}
consistent with \citet{nenkova08-pt2}.  Here $\xi_{\rm dust}$ is a free parameter
indicating the fraction of disk photons that are reprocessed by the
dust torus.  For the BLR photons, 
\begin{equation}
u_* = u_{\rm line} = \frac{u_{\rm line,0}}{1 + (r_{\rm blob}/r_{\rm line})^\beta} \ ,
\label{eq-uline}
\end{equation}
where $\beta\approx 7.7$ \citep{finke16}, $r_{\rm blob}$ is the
distance of the emitting blob from the black hole (a free parameter).
The line radii $r_{\rm line}$ and initial energy densities $u_{\rm
  line,0}$ for all broad lines used are given by the Appendix of
\citet{finke16} relative to the H$\beta$ line based on the composite
SDSS quasar spectrum of \citet{vanden01}.  The parameters $r_{\rm
  line}$ and $u_{\rm line,0}$ are determined from the disk luminosity
using relations found from reverberation mapping, as described by
\citet{finke16} and in Paper~2.

\subsection{High-Energy Attenuation}
\label{ggabsorb}

Since the analysis of the 2013 December 20 flare predicts very high
energy $\g$-rays, which could be attenuated, we include
$\g\g$-absorption from the dust torus and BLR photons following
\citet{finke16}.  Absorption attenuates the emerging jet emission by a
factor $\exp[-\tau_{\g\g}(\epsilon_1)]$ where $\tau_{\g\g}(\e_1)$ is
the absorption optical depth and $\epsilon_1$ is the dimensionless
energy of the higher energy photon produced in the jet.  We have computed the Doppler factor where $\g\g$ absorption with internal synchrotron photons becomes important \citep[e.g.][]{dondi95,finke08_SSC} and found that the minimum Doppler factor is much lower than the value used in our models here (Table \ref{tbl-freeparams}).  Therefore, we hereafter neglect internal synchrotron photoabsorption.

Emission from the BLR comes from a relatively narrow region at
sub-parsec scales from the BH and different lines are produced at
different radii \citep[e.g.][]{peterson99,kollatschny03,peterson14},
which can consist of concentric, infinitesimally thin, spherical
shells for each emission line or concentric, infinitesimally thin
rings Similarly, the dust torus can be modeled as a ring with an
infinitesimally thin annulus or a more extended flattened disk with
defined inner and outer radii. After testing each geometry, we find
that in all cases $\g\g$-absorption is unimportant to model the energy
range studied here, although attenuation by dust torus photons can
have some effect at $\ga800$\ GeV.  The following analysis utilizes the 
concentric shell BLR and ring dust torus geometries, which are consistent 
with the emission calculations. 

\section{Application to 3C 279: 2013 December Flare and Preceding Quiescent Period}
\label{3c279}

The three days immediately preceding the extreme, Compton-dominant
flare of the FSRQ, 3C 279, were quiescent and apparently unremarkable
for the source.  This period, dubbed ``Epoch A'' in \citet{hayash15},
where the data were originally published occurred on 2013 December
16-19, and is analyzed to look for any unusual parameter values and to
provide context for the flare analysis.  The isolated $\g$-ray flare
occurred during a 12 hour period on 2013 December 20, dubbed ``Epoch
B'' by \citet{hayash15}.  In both Epochs A and B, besides the $\g$-ray
data from {\em Fermi}-LAT, optical and IR data were collected by the
Katana Telescope and SMARTS.  During Epoch A, radio observations were
made by the Sub-Millimeter Array, UV and optical data were collected
by {\em Swift}-UVOT, and X-ray data were collected by both {\em
  NuSTAR} and {\em Swift}-XRT.  Due to the unexpected nature and short
duration of the flare (Epoch B), there were no radio or X-ray
observations during that time.  We analyzed these SEDs using the model
described in Section \ref{model} with the full Compton expressions for radiative
cooling and emission, and the numerical solution to the Fokker-Planck
equation.

\subsection{Particle Acceleration and Spectral Emission}

\begin{figure}
\vspace{2.2mm} 
\centering
\includegraphics[width=0.5\textwidth]{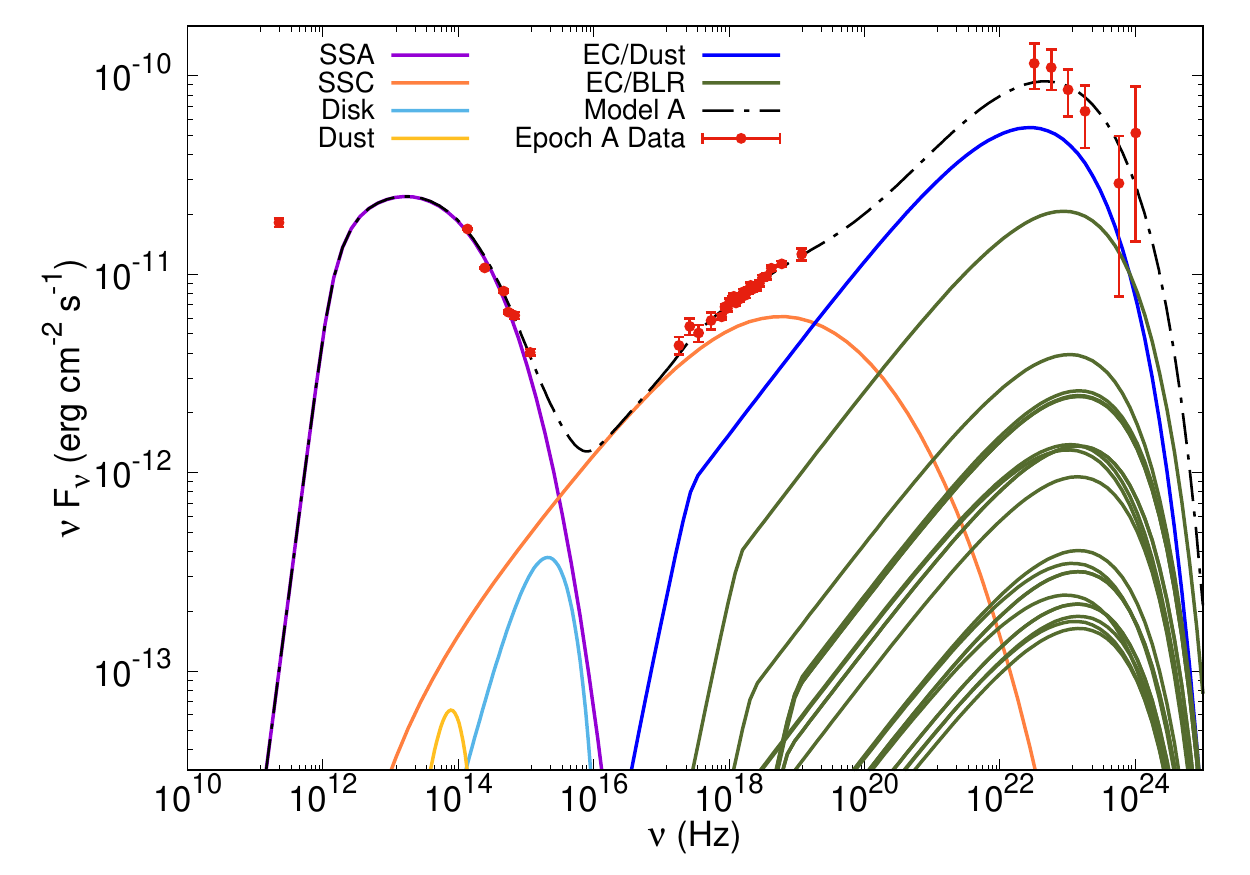}
\caption{The SED of 3C 279 during Epoch A \citep[2013 December
    16-19;][]{hayash15} and our model results.  Curves are labeled in
  the legend.}
\label{fig-Aemiss}
\vspace{2.2mm}
\end{figure}

\begin{deluxetable}{llllr}
\tablecaption{Free Model Parameters \label{tbl-freeparams}}
\tablehead{
\colhead{Parameter (Unit)}
& \colhead{Model A}
& \colhead{Model B1}
& \colhead{Model B2}
& \colhead{Model B3}
}

\startdata
$t_{\rm var}$ (s)
&$1.5 \times 10^{4}$
&$3.5 \times 10^{4}$
&$2.3 \times 10^{3}$
&$9.0 \times 10^{2}$
\\
$B$ (G)
&$1.3$
&$0.07$
&$0.21$
&$0.3$
\\
$\delta_{\rm D}$ 
&$30$
&$18$
&$49$
&$70$
\\
$r_{\rm blob}$ (cm)
&$1.9 \times 10^{17}$
&$1.4 \times 10^{17}$
&$1.6 \times 10^{17}$
&$1.3 \times 10^{17}$
\\
$\xi_{\rm dust}^{\dagger}$ 
&$0.1$
&$0.1$
&$0.1$
&$0.1$
\\
$T_{\rm dust}^{\dagger}$ (K)
&$1470$ 
&$1470$ 
&$1470$ 
&$1470$ 
\\
$L_{\rm disk}^{\dagger}$ (erg s$^{-1}$)
&$1.0 \times 10^{45}$ 
&$1.0 \times 10^{45}$ 
&$1.0 \times 10^{45}$ 
&$1.0 \times 10^{45}$ 
\\
$D_0$ (s$^{-1}$)
&$7.0\times 10^{-6}$
&$2.5\times 10^{-6}$
&$9.0\times 10^{-6}$
&$1.5\times 10^{-5}$
\\
$a$
&$-4.1$
&$+1.0$
&$-2.0$
&$-0.5$
\\
$\gamma_{\rm inj}^{\dagger}$
&$1.01$
&$1.01$
&$1.01$
&$2.01^{\ddagger}$
\\
$L_{\rm inj}$ (erg s$^{-1}$)
&$8.8 \times 10^{28}$
&$1.0 \times 10^{32}$
&$8.5 \times 10^{30}$
&$5.4\times 10^{30}$
\enddata
\tablenotetext{\dagger}{Parameter held constant during analysis, although implemented as free.}
\tablenotetext{\ddagger}{The positive slope in the ED at $\g \lesssim 10$ confounds the numerical normalization scheme, but the results are essentially equivalent to a Model B3 with $\gamma_{\rm inj}=1.01$ due to low particle injection number. }
\end{deluxetable}

We first analyzed Epoch A with our model.  The $\nu F_{\nu}$ SED data
for Epoch A is shown in Figure \ref{fig-Aemiss} with our model result,
and our model parameters are in Table \ref{tbl-freeparams}.  Figure
\ref{fig-Aemiss} demonstrates that the $\g$-ray data is described by a
27 component EC model, including the dust torus and a stratified BLR.  Scattering of dust torus photons is the largest single contributor to the production of $\g$-ray emission, although the sum of all BLR photon scattering is similar.
The X-ray data are predominantly reproduced by the SSC process
although EC/Dust contributes heavily at harder energies.  The IR-UV
data is reproduced primarily by synchrotron radiation, which includes
self-absorption at lower energies.  Note that the radio emission is
produced outside of the modeled region, so that the data are upper
limits on the model.  


\begin{figure}
\vspace{2.2mm} 
\centering
\includegraphics[width=0.5\textwidth]{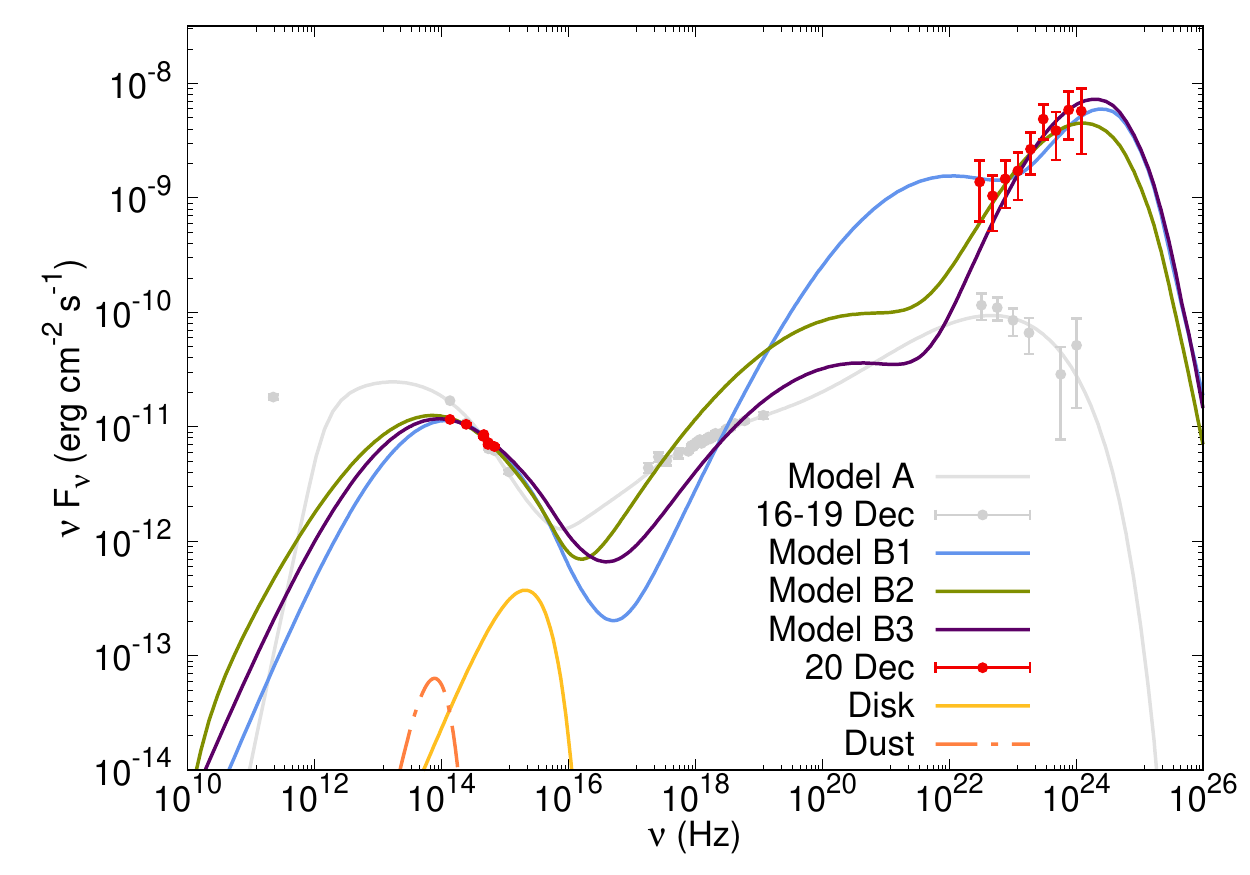}
\caption{The SED of 3C 279 during Epoch B \citep[2013 December
    20;][]{hayash15} and our 3 model results (models B1, B2, and B3).
  Curves are labeled in the legend.  The data and model for Epoch A
  are also shown in light grey for reference.  }
\label{fig-Bcomposite}
\vspace{2.2mm}
\end{figure}

Figure \ref{fig-Bcomposite} shows spectral data for Epoch B, which was
observed during a 12 hour period on 2013 December 20, immediately
following Epoch A.  There are no X-ray data available during the 12
hours of peak flare for Epoch B, on 2013 December 20, and as shown in
the analysis of Epoch A (Figure \ref{fig-Aemiss}), the X-rays
constrain primarily the SSC component of the spectrum.  Thus, the
variability timescale $t_{\rm var}$ and Doppler beaming factor $\dD$
are not well constrained for this epoch.  However, since the electron
distribution informs the SED and the parameters for the SED emission
are all included in the ED as loss parameters, none
of the jet components in the spectral model are fully independent from
one another.  We explore the parameter space with three models, each
making different predictions for X-ray emission.  Our model parameters
for each one are in Table \ref{tbl-freeparams}. Model B1 is the only
one to reproduce the ``ankle'' in the $\g$-ray spectrum (i.e., the
change in spectral index around 300 MeV), but requires a high SSC flux
not previously observed for this source, leading to a very hard X-ray
spectrum.  For all of the Epoch B models, the $\g$-ray emission is
dominated by scattering of BLR photons (unlike our model for Epoch A,
where it was dominated by the scattering of dust photons).  Model B2
is an intermediate possibility and predicts a more moderate X-ray
spectrum.  Model B3 has both the lowest flux and the lowest frequency
peak for the SSC.  Model B3 also has the smallest timescale for
particle acceleration (Section \ref{timescales}), which is important since the flux-doubling
timescale for the flare was quite short ($\sim \, 2\,$hr).  Model B3
has a high Doppler factor, indicating a very high blob velocity.  All
of the models for Epoch B have different $\dD$ (and therefore $\G$)
from the Epoch A model.  However, this is not a problem, as the
emission in Epochs A and B are likely produced by different blobs,
which can be moving with different $\G$, since within a few days for
the observer, the distance of a given emitting blob from the BH
$r_{\rm blob}$ will change significantly.  In Figure
\ref{fig-Bcomposite} the results of the Epoch A analysis are also
shown for reference, demonstrating both the difference in the observed
$\g$-ray spectra, as well as the possible changes to the X-rays.  All
of the B models have harder X-ray spectra for the flare than was
observed during the quiescent period, which is qualitatively
consistent with the optical and $\g$-ray hardness changes.
\citet{hayash15} estimate the Compton dominance $A_{\rm C} \ge 300$
for Epoch B, and thus classify the flare as a rare, extreme Compton
flare. We predict Compton dominance values up to  $A_{\rm C} = 1800$
for Model B3 (Table \ref{tbl-calculated}).

Throughout the simulations, we hold constant several parameters that the
model can in principle vary (see Table \ref{tbl-freeparams}).  The
parameters $L_{\rm disk}$, $\xi_{\rm dust}$, and $T_{\rm dust}$ are
not expected to vary significantly on timescales of days.  The Lorentz
factor of particles injected into the base of the blob $\g_{\rm inj}$
is generally taken to be near unity as these particles are expected to
originate from the thermal population in the accretion disk.  However
the numerical machinery can produce incomplete solutions for positive
ED slopes near $\g_{\rm inj}$ when $\g_{\rm inj}/\g_{\rm max}$ is very
small (and for negative ED slopes near $\g_{\rm inj}$ as $\g_{\rm
  inj}/\g_{\rm max}$ goes to 1).  Thus, for Model B3, we used $\g_{\rm
  inj}=2$.  However,
since the  number of particles injected into the blob is low compared to the total
number of particles in the ED at the injection energy, the ED solution
is effectively the same.


Models A, B2, and B3 have stronger second-order Fermi acceleration (as
indicated by a larger $D_0$), while in Model B1 first-order Fermi process
dominates over adiabatic cooling ($A_{\rm ad+sh}>0$).  Notably, all of the
models are able to represent the available broadband
multiwavelength data, suggesting that it is possible for Fermi
acceleration processes to meet the acceleration requirements for the
observed emission.

In Figure \ref{fig-Bcomposite}, the data suggest that the peak luminosity of the
synchrotron emission decreases during the flare, which might indicate
a decrease in the magnetic field $B$ for a similar particle distribution.  In
Model B1, this effect is exaggerated, with the particle distribution being much more energetic than in
Model A, and thus the magnetic field strength is especially low (see Table
\ref{tbl-freeparams}).


The dust and BLR energy densities ($u_{\rm dust}$ and $u_{\rm BLR}$, respectively) represent the energy in the incident photon fields  in the AGN rest frame, and that are available for external Compton (EC) scattering.  
The combined BLR energy density is elevated by a factor of a few in Model B1 from the Model A values, which are within the previously observed range.  The elevated energy density of EC photon sources should be expected of an orphaned $\g$-ray flare for a similar ED, since the Compton dominance $A_{\rm C}$ is high (see Table \ref{tbl-calculated}).  However, in the case of Model B1, the requisite external energy densities are lower because more of the scattering energy is provided by the elevated energy in the ED.  Thus Model B1 has more moderate energy requirements from the environment outside the jet than Models B2 and B3.   

\begin{deluxetable}{llllllr}
\tablecaption{Calculated Parameters \label{tbl-calculated}}
\tablewidth{0pt}
\tablehead{
\colhead{Parameter (Unit)}
& \colhead{Model A}
& \colhead{Model B1}
& \colhead{Model B2}
& \colhead{Model B3}
}

\startdata 
$R_{Ly\alpha}$ (cm)
&$2.7 \times 10^{16}$ 
&$2.7 \times 10^{16}$  
&$2.7 \times 10^{16}$ 
&$2.7 \times 10^{16}$ 
\\
$R_{H\beta}$ (cm)
&$1.0 \times 10^{17}$ 
&$1.0 \times 10^{17}$ 
&$1.0 \times 10^{17}$ 
&$1.0 \times 10^{17}$ 
\\
$R\p_{b}$ (cm)
&$8.5 \times 10^{15}$
&$1.2 \times 10^{16}$
&$2.4 \times 10^{15}$
&$1.2 \times 10^{15}$
\\
$\phi_{j,{\rm min}}$ ($^{\circ}$)
&$1.3$
&$2.4$
&$0.4$
&$0.5$
\\
\hline
\\
$P_B$ (erg s$^{-1}$)
&$8.2 \times 10^{44}$
&$1.6 \times 10^{42}$
&$4.5 \times 10^{42}$
&$5.0 \times 10^{42}$
\\
$P_e$ (erg s$^{-1}$)
&$8.6 \times 10^{45}$
&$2.2 \times 10^{46}$
&$9.0 \times 10^{45}$
&$4.3 \times 10^{45}$
\\
$\zeta_e^{\dagger}$
&$1.0 \times 10^{1}$
&$1.4 \times 10^{4}$
&$2.0 \times 10^{3}$
&$8.5 \times 10^{2}$
\\
$\mu_{a}^{\ddagger}$
&$3.8$
&$8.8$
&$3.6$
&$1.7$
\\
\hline
\\
$u_{\rm ext}$ (erg cm$^{-3}$)
&$3.5\times 10^{-4}$
&$9.6\times 10^{-4}$
&$5.8\times 10^{-4}$
&$1.3\times 10^{-3}$
\\
$u_{\rm dust}$ (erg cm$^{-3}$)
&$1.6 \times 10^{-4}$ 
&$1.6 \times 10^{-4}$ 
&$1.6 \times 10^{-4}$ 
&$1.6 \times 10^{-4}$ 
\\
$u_{\rm BLR}$ (erg cm$^{-3}$)
&$1.9 \times 10^{-4}$
&$8.0 \times 10^{-4}$
&$4.2 \times 10^{-4}$
&$1.1 \times 10^{-3}$
\\
\hline
\\
$A_C$
&$4.7$
&$1600$
&$800$
&$1800$
\\
$L_{\rm jet}$ (erg s$^{-1}$)
&$5.0 \times 10^{44}$
&$5.4 \times 10^{46}$
&$4.5 \times 10^{45}$
&$3.1 \times 10^{45}$
\\
$\sigma_{\rm max}$
&$4.3$
&$2.9$
&$2.2$
&$1.8$
\\
\enddata
\tablenotetext{\dagger}{ $\zeta_e = u_e/u_B = P_e/P_B$ is the equipartition parameter, where $\zeta_e = 1$ indicates equipartition.}
\tablenotetext{\ddagger}{ $\mu_a \equiv (P_B+P_e)/P_a$ is the ratio of jet to accretion power.  Magnetically arrested accretion explains values of $\mu_a \lesssim$ a few. } 
\end{deluxetable}



Models B2 and B3 employ more moderate estimates of the SSC flux, and the magnitude and spectral index in the X-rays are closer to those observed during Epoch A (Figure \ref{fig-Bcomposite}).  In both Models B2 and B3, the magnetic field $B\lesssim0.3\,$G (Table \ref{tbl-freeparams}), is consistent with the analysis in \citet{hayash15}. These higher (than Model B1) values for the magnetic field strength produce similar synchrotron simulations because in Models B2 and B3, there is less energy in the jet electrons (Figure \ref{fig-ElecDist}) and a lower $\dD$.  Additionally, the lower energy ED requires higher $\dD$ and higher energy densities of the external radiation fields due to dust and the BLR to maintain the Compton dominance in the EC portion of the simulation. 


\subsection{The Electron Distribution}

\begin{figure}[t]
\vspace{2.2mm} 
\centering
\includegraphics[width=0.5\textwidth]{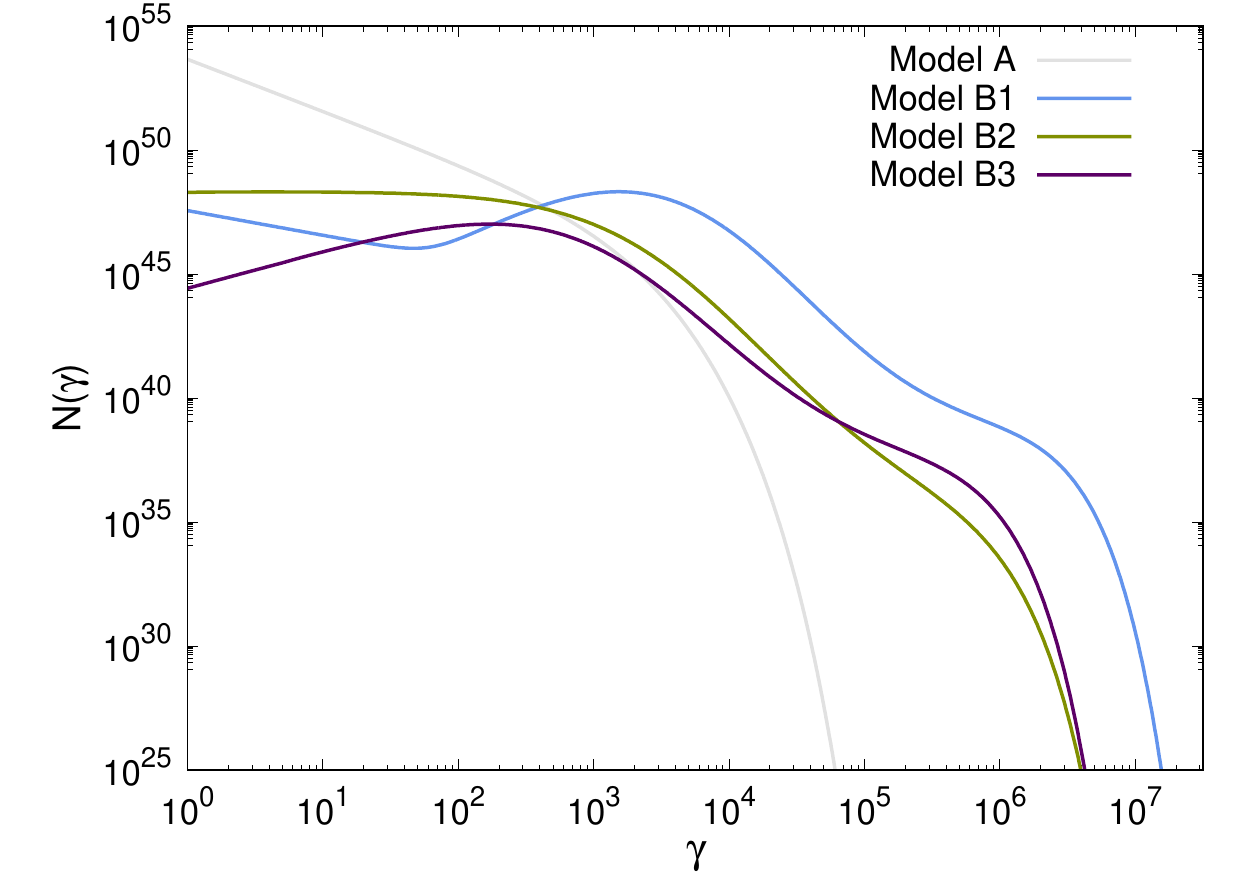}
\caption{The numerical EDs for the model SEDs in Figure \ref{fig-Bcomposite}.  Models B2 and B3 have more particles at higher energies than Model A.  Similarly, Model B1 has the most particles occupying the highest energies.}
\label{fig-ElecDist}
\vspace{2.2mm}
\end{figure}

Since the ED for each SED is produced independently from the others, and is an integral part of the simulation, it is instructive to look at the shapes produced for each model.  Figure \ref{fig-ElecDist} has ED curves for each SED model in Figure \ref{fig-Bcomposite}, with the same color scheme.  Model A (grey) appears as a relatively simple power-law with an exponential cutoff, which is due to the balance between  first-order Fermi (including adiabatic expansion) and second-order Fermi accelerations, where the ratio between the two $a=-4.1< a_{\rm critical} = -2$, thus $N(\g)$ is decreasing with increasing $\g$.  The second-order Fermi acceleration dominated portion of the solution is subdominant in Epoch A (acceleration dependencies in the ED are derived in Appendix \ref{ap-deps}).

In each of the Epoch B model EDs, both terms from Equation (\ref{eq-anasim}) are apparent, but they are arranged differently in Figure \ref{fig-ElecDist} than in Figure \ref{fig-ElecDist_Simp}.  The second-order Fermi term, which was cut off around $\g_{\rm max} \sim 10^5$ in Figure \ref{fig-ElecDist_Simp} is cut off at $\g_{\rm max} \sim 10^{7}$ for Model B1 and $\g_{\rm max} \sim 10^{6.5}$ for Models B2 and B3 (Figure \ref{fig-ElecDist}).  It is interesting to note that for Model B1, the second-order Fermi component dominates at $10^2\la \g \la10^5$.  All of the cut-offs occur at much higher Lorentz factors than Model A, indicating that acceleration is providing more energy to the particles.  So, regardless of the particular simulation, the flare requires more high-energy particles than does the preceding quiescent period. The feature in the EDs at $\g\approx 10^{3.5}$ is due to the first-order Fermi/adiabatic/second-order Fermi term in Equation (\ref{eq-anasim}).  Model B2 has $a=-2$, which produces a slope of $0$ in that term.  Models B1 and B3 have $a>-2$, and produce positive slopes at $\g\lesssim10^{3.5}$ in the balanced term. Positive slopes indicate that acceleration outpaces emission in that energy range.  However, the combined term acceleration is damped by emission mechanisms at much lower energies than the second-order Fermi dominated term for all three Epoch B models.

\begin{figure}[t]
\vspace{2.2mm} 
\centering
\includegraphics[width=0.5\textwidth]{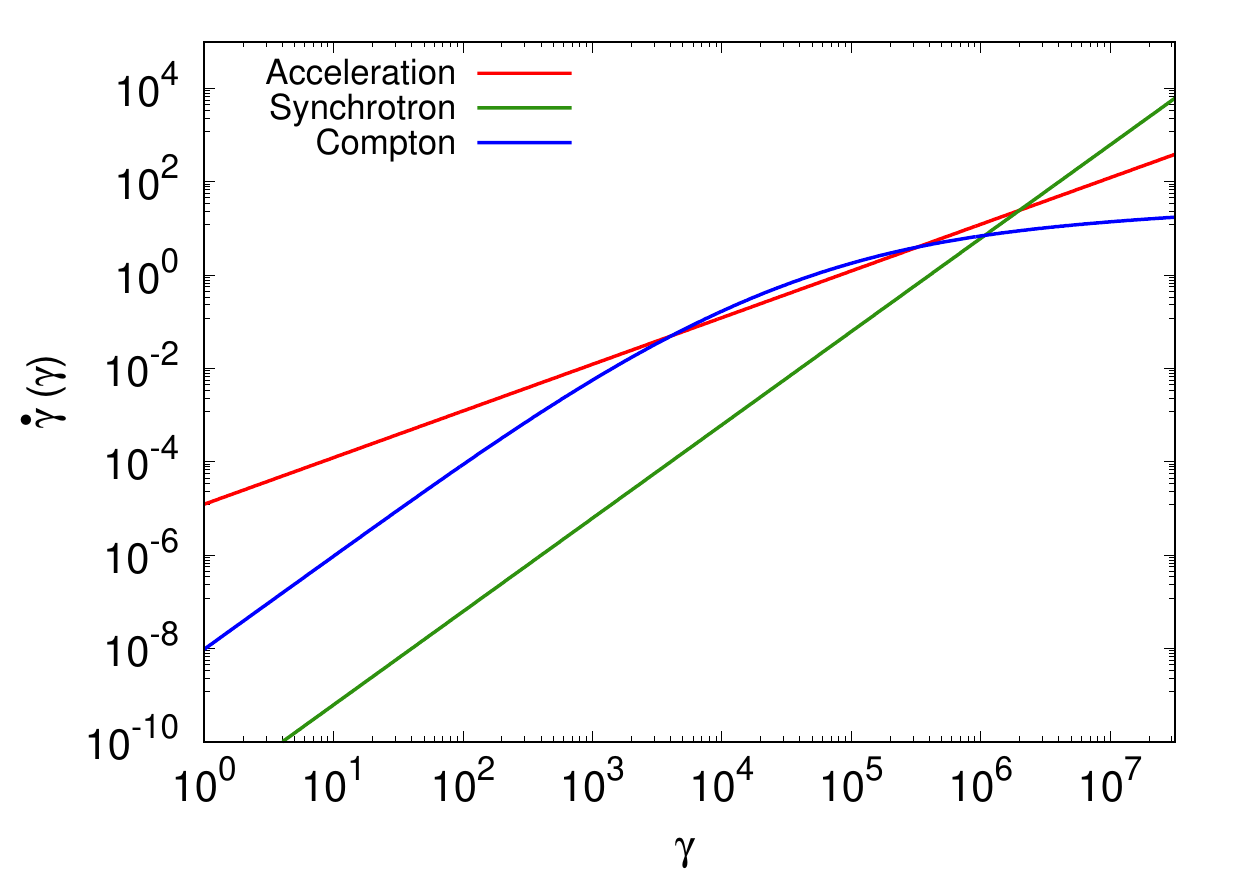}
\caption{The electron acceleration rate and synchrotron and Compton scattering radiative loss rates in the comoving frame for Model B1.} 
\label{fig-gammaDot}
\vspace{2.2mm}
\end{figure}

The approximation Equation (\ref{eq-anasim}) is based on the analytic derivation in the Thomson limit.  In the numerical solution, we include the full Compton cross-section, and it is possible to separate the loss mechanisms.  Figure \ref{fig-gammaDot} shows the rate of energy gain and loss at each Lorentz factor for Model B1.  The red acceleration curve in Figure \ref{fig-gammaDot} is given by the sum of acceleration rates for first- and second-order Fermi processes (Equation [\ref{eq-AdShRate}] and Equation [\ref{eq-StoRate}], respectively).  Acceleration dominates at Lorentz factors $\g \lesssim 10^{3.5}$, where it is intersected by the blue Compton curve, which corresponds to the first turnover in the ED (Figure \ref{fig-ElecDist}).  The Compton loss rate (Equation [\ref{eq-CompRate}]) changes at higher energies due to the Klein-Nishina effect, which allows it to trace the acceleration rate through $\g\sim 10^5$, causing a decline in the ED over the same range (Figure \ref{fig-ElecDist}).  The synchrotron loss rate (Equation [\ref{eq-SynRate}]) is subdominant until $\g \gtrsim10^6$, at which point it provides the definitive exponential cutoff to the ED.

\subsection{Jet Dynamics and Geometry}
\label{sec-dyngeo}

Our model gives $r_{\rm blob} \sim (5-7) \times 10^{17}$\ cm during
previous quiescent and flare states of 3C 279 (Paper~2).  If the X-ray emission does not
significantly change between Epochs A and B, then the emitting region
size $R\p_b$ may decrease as acceleration and emission increase.  In
all our models the minimum jet opening angle has a reasonable value.

It is clear within our analysis that first- and second-order Fermi accelerations are 
sufficient to power the observed flare, even assuming that the source 
particles are from a thermal distribution. Thus, no other forms of 
acceleration are necessary to explain the behavior. However, we briefly 
explore reconnection because it is part of the existing conversation in 
the literature of this particular flare's acceleration.  Magnetic 
reconnection can in principle describe a rapid flare with very short variability timescale
and a hard particle spectrum, both of which are observed.  It requires a
magnetization parameter, $\sigma\gg1$.  This can be constrained by (Paper~1)
\begin{flalign}
\sigma < \sigma_{\rm max} = \frac{3D_0R\p_b}{c}\ ,
\end{flalign}
based on the maximum Larmor radius of an electron that fits inside the 
blob. The model calculations of $\sigma_{\rm max}$ can be found in Table
\ref{tbl-calculated}.  In all models, $\sigma_{max}\sim 1$.  Since particle 
reconnection generally requires $\sigma \gg 1$ , a lack of particle 
acceleration by reconnection is consistent with our models.  First- and second-order Fermi 
acceleration are sufficient to explain the emission during the 
epochs explored here.

Equipartition is used in a wide range of astrophysical analyses, and
can be used as a simplifying assumption for otherwise poorly
constrained parameters \citep[e.g.][]{dermer14}.  There is an
uncertainty involved, since the power in protons (either accelerated
or ``cold'') is poorly constrained from the modeling
\citep[e.g.][]{beck05}.  In the discussion in the rest of this
section, we neglect the presence of protons in the jet; however, this
uncertainty should be kept in mind.  In our models for 3C 279 we find
the power in the field $P_B$ is not always equivalent to the
power in the jet electrons $P_e$, although $P_B\sim P_e$ within an
order of magnitude or two for previously examined epochs (Paper~2).
Our Model A analysis shows the quiescent jet was close to equipartition
(Table \ref{tbl-calculated}), where the equipartition parameter
$\zeta_e \equiv P_e/P_B =1$ represents an equipartition state.  However, in
all of the B models $P_B\ll P_e$ (Table \ref{tbl-calculated}), and the
jet is strongly electron dominated. The larger the frequency
integrated flux, the larger the equipartition parameter for Epoch B.
This is consistent with the analysis by \citet{hayash15} for the Epoch B
flare; they also found the jet was matter dominated.  A matter
dominant jet might be indicative of a larger than usual influx of
electrons into the emitting region. An analysis for Epoch B was
attempted, in which $P_B\approx P_e$
but the jet opening angle was unphysically
large, and therefore it is not presented here.

Neglecting protons, the total jet power $P_{B+e}=P_e + P_B$.  For a
maximally rotating BH, one expects that the accretion power $P_{\rm a}
= L_{\rm disk}/0.4$, giving $P_{\rm a} =
2.4\times10^{45}\ \erg\ \s^{-1}$ for our models.  For all our models
except model B1, $\mu_a \equiv P_{B+e}/P_a$ is a factor $\sim$\ a few
(Table \ref{tbl-calculated}).  This is consistent with extracting spin
from a black hole with a magnetically arrested accretion disk
\citep{tchekhovskoy11}.  However, for Model B1, $\mu_a=8.8$, which
may be too large too extract from the \citet{blandford77} process.

The total jet luminosity due to particle radiation \citep[e.g.][]{finke08_SSC}
\begin{flalign}
L_{\rm jet} = \frac{2\pi d_L^2}{\G^2}\int_0^\infty \frac{d\e}{\e}\ 
\left(f_{\epsilon_{\rm obs}}^{\rm syn} + f_{\e_s}^{\rm SSC} + \sum_{j=1}^J f_{\e_s}^{\rm EC,(j)}\right)\\
\vspace{2mm} \nonumber
\end{flalign}
is reported in Table \ref{tbl-calculated}.  The radiative efficiency
$L_{\rm jet}/P_{B+e}$ is expected to be $<1$.  This is indeed the
case, for all models except for Model B1.

Based on the excessive $\mu_a$ and large radiative efficiency ($L_{\rm
  jet}/P_{B+e}>1$), we believe Model B1 is unphysical and cannot
explain Epoch B.

The injection luminosity $L_{\rm inj}$ of Epoch A is similar to other
injection luminosities we have found for 3C 279 in other epochs
(Paper~2).  The injection luminosities for the B models are somewhat
higher due primarily to the increased rate of particle injection.
This supports the idea of an influx of particles from the accretion
disk area instigating the flare.

\subsection{Energy Budget}
\label{budget}

One benefit of the Fokker-Planck equation (or transport equation)
formalism is the conservation of energy.  We can compute the rate of
energy gains and losses in electrons (in the co-moving frame) that is: injected into
the system ($P\p_{\rm inj}$); escaping the system ($P\p_{\rm
  esc}$); accelerated by the second-order Fermi acceleration process
($P\p_{\rm sto}$); accelerated by first-order Fermi acceleration or lost
by adiabatic expansion ($P\p_{\rm sh+ad}$); lost by
synchrotron ($P\p_{\rm syn}$) or EC radiation ($P\p_{\rm EC}$).  The
first-order Fermi acceleration and adiabatic losses are taken together, since in practice 
separating them introduces an unconstrainable free parameter in our formalism.  
(See Paper~2 for the details on how these rates are calculated.) 

The component powers resulting from our simulations for the Epochs examined here can be found in Table
\ref{tbl-powers}.  The percent relative errors 
found by adding up the powers are low (within expected numerical
errors) indicating the expected energy balance.

Since in all cases $P\p_{\rm sto}>P\p_{\rm sh+ad}$, the particles are
primarily accelerated by the second-order Fermi process, rather than first-order Fermi
acceleration.  This is consistent with our results for other SEDs in Paper~1 and Paper~2.  First-order Fermi acceleration dominates over
adiabatic losses for only model B1, since this is the only simulation where
$P\p_{\rm sh+ad}>1$.  This is consistent with this being the only
model with $A_{\rm ad+sh}>0$ (Table \ref{tbl-freeparams}).  For the other models,
adiabatic losses are an important energy loss mechanism.  For Model A,
adiabatic losses dominate over radiative losses by a factor of $\sim
100$.  For models B2 and B3, they are the same order of magnitude as
radiative losses.  We note the Epoch B models have a much larger
$P\p_{\rm EC}/P\p_{\rm syn}$ than the Epoch A simulation, as expected,
since Epoch B has a much larger $A_C$ than Epoch A.

The injection energy is not an important contribution to the energy
budget.  The escape power is always slightly larger in magnitude than
the injection power because the escape occurs in the Bohm limit,
meaning higher energy particles are preferentially lost from the
electron population (Paper~1).

\begin{deluxetable}{llllllr}
\tablecaption{Power in the Physical Components \label{tbl-powers}}
\tablewidth{0pt}
\tablehead{
\colhead{Variable (Unit)}
& \colhead{Model A}
& \colhead{Model B1}
& \colhead{Model B2}
& \colhead{Model B3}
}

\startdata
$P\p_{\rm esc}$ (erg s$^{-1}$)
&$-6.0 \times 10^{31}$
&$-5.3 \times 10^{35}$
&$-9.6 \times 10^{33}$
&$-2.9 \times 10^{33}$
\\
$P\p_{\rm inj}$ (erg s$^{-1}$)
&$8.8 \times 10^{29}$
&$1.0 \times 10^{32}$
&$8.6 \times 10^{30}$
&$1.3 \times 10^{31}$
\\[5pt]
$P\p_{\rm sto}$ (erg s$^{-1}$)
&$3.2 \times 10^{43}$
&$1.9 \times 10^{44}$
&$7.2 \times 10^{42}$
&$1.4\times 10^{42}$
\\
$P\p_{\rm sh+ad}$ (erg s$^{-1}$)
&$-2.4 \times 10^{43}$
&$+4.7 \times 10^{43}$
&$-3.6 \times 10^{42}$
&$-1.8 \times 10^{41}$
\\
$P\p_{\rm syn}$ (erg s$^{-1}$)
&$-2.4 \times 10^{41}$
&$-6.3 \times 10^{41}$
&$-1.3 \times 10^{40}$
&$-3.0 \times 10^{39}$
\\
$P\p_{\rm EC}$ (erg s$^{-1}$)
&$-8.6 \times 10^{41}$
&$-2.4 \times 10^{44}$
&$-3.6 \times 10^{42}$
&$-1.3 \times 10^{42}$
\\
$\%\sigma_{\rm err}$
&$2.9$
&$0.03$
&$0.03$
&$0.04$
\enddata
\end{deluxetable}

\subsection{Acceleration Timescales}
\label{timescales}

An additional benefit of the particle transport method we employ is the ability to compare the timescales for each physical process, as expressed by the coefficients in Equation [\ref{eq-StoRate}] through Equation [\ref{eq-CompRate}], which have units of [s$^{-1}$] (Paper~1). Of particular interest during the flare is the acceleration timescale, because the flare duration is short ($\sim 12\,$hr) and the flux-doubling timescale is rapid ($\sim 2\,$hr). Hence the acceleration mechanism providing energy to the flare must act on commensurate timescales \citep[e.g.][]{hayash15,paliya16}.

\begin{deluxetable}{llllllr}
\tablecaption{Simulated Emission \& Acceleration Timescales \label{tbl-times}}
\tablewidth{0pt}
\tablehead{
\colhead{Variable (Unit)}
& \colhead{Model A}
& \colhead{Model B1}
& \colhead{Model B2}
& \colhead{Model B3}
}

\startdata
$t_{\rm sto}$ (h)
&$0.7$
&$2.5$
&$0.2$
&$0.1$
\\
$t_{\rm sho}$ (h)
&$--$
&$\lesssim\,1.1$
&$\lesssim\, 3.5$
&$\lesssim\, 0.5$
\\[5pt]
$t_{\rm acc}$ (h)
&$\le 0.7$
&$\lesssim 0.7$
&$\lesssim 0.2$
&$\lesssim 0.1$
\enddata
\end{deluxetable}

The mean timescale for the second-order Fermi acceleration of electrons via hard-sphere scattering with MHD waves is computed in the frame of the observer using
\begin{equation}
t_{\rm sto} = \frac{1+z}{\dD}\frac{1}{4D_0}  \ ,
\end{equation}
where the factor of 4 comes from the derivative separating the broadening and drift coefficient components of the second-order Fermi up-scattering.
First-order Fermi acceleration is closely linked with adiabatic expansion in the transport model during analysis of the SED, since both processes have the same energy dependence (Paper~1). However, we constrain the first-order Fermi acceleration timescale by assuming that the energy loss rate due to adiabatic expansion during the flare, $A_{\rm ad}^{\rm flare}$, is the same as during quiescence, $A_{\rm ad}^{\rm quiescent}$.
Thus, the first-order Fermi acceleration timescale during the flare can be constrained using Model A as the limiting case, which yields
\begin{equation}
t_{\rm sh} \lesssim \frac{1+z}{\dD}\frac{1}{A_{\rm ad + sh}^{\rm flare} - A_{\rm ad}^{\rm quiescent}}  \ ,
\label{tshock}
\end{equation}
where $A_{\rm ad}^{\rm quiescent} \lesssim A_{\rm sh+ad}^{\rm quiescent}$ (values in Table \ref{tbl-freeparams}).  
The total acceleration timescale, $t_{\rm acc}$, depends on both the first- and second-order Fermi timescales, $t_{\rm sh}$ and $t_{\rm sto}$, respectively, via
\begin{equation}
t_{\rm acc} = \frac{1}{t_{\rm sh}^{-1} + t_{\rm sto}^{-1}} \ ,
\end{equation}
where the value of $t_{\rm sh}$ is an upper limit given by Equation~(\ref{tshock}).  The values provided in Table \ref{tbl-times} for $t_{\rm acc}$ are therefore also upper limits.  
%
Thus, first- and second-order Fermi acceleration rates, as included in the model, are sufficiently rapid to produce the observed flux doubling timescale for the 2013 flare.

\section{Discussion}
\label{discuss}

In the following section, we discuss the physical implications of the analysis, compare our results with previous literature, and summarize our primary findings. 

\subsection{Physical Interpretations}

Blazar jets are thought to contain standing shocks \citep[e.g.][]{marscher08}.  A superluminal blob passing through a standing shock will increase the amount of first-order Fermi acceleration \citep[e.g.,][]{marscher12}, which is consistent with the Epoch B models, compared with our model for Epoch A (Table \ref{tbl-freeparams}).  MHD simulations demonstrate that first-order Fermi acceleration gives rise to higher levels of stochastic turbulence downstream of the shock \citep[e.g.][]{inoue11}, which agrees with the increase in second-order Fermi acceleration during the flare analyses (Table \ref{tbl-freeparams}).  Both the first- and second-order Fermi acceleration contribute to the higher maximum electron Lorentz factor as well as the higher number of high-energy particles (see Figure \ref{fig-ElecDist}).  These are important to the higher frequency position of each emission component in the SED. 

As particle energy increases, so does the Larmor radius ($r_{\rm L} = \g m_e c^2 q^{-1} B^{-1}$).  A particle with a larger Larmor radius, will travel preferentially closer to the edge or sheath of the jet \citep[e.g.][]{hillas84}.  If the magnetic field is radially dependent (stronger near the jet core), then the apparent magnetic field of the jet may be lower by a factor of a few when more particles spend more time near the edge of the emitting region in the jet.  \citet{massaro04} discuss a log-parabolic particle distribution (which can be formed by second-order Fermi acceleration; \citealt{tramacere11}) as one in which the confinement efficiency of a collimating magnetic field decreases with increasing gyro-radius.  This physical interpretation can explain the smaller magnetic field (Table \ref{tbl-freeparams}) and the greater loss to escaping particles (Table \ref{tbl-powers}), during the flare.  

In Model B1, the first-order Fermi acceleration contributes more than in Models B2 and B3.  This is consistent with the smaller bulk Lorentz factor (recall we assume $\dD=\G$, Table \ref{tbl-freeparams}), since increased first-order Fermi acceleration indicates a given particle is scattered through the shock front more times.  Thus, even larger Larmor radii may indicate that emitting particles occupy some of the jet sheath, where the magnetic field is significantly lower, which explains the $95\%$ drop in field strength in the analysis of the flare (Table \ref{tbl-freeparams}).  While the interpretation is somewhat more extreme for Model B1, the second-order Fermi acceleration coefficient and variability timescale are more similar to Model A than are  Models B2 and B3.  However, Model B1 has an unphysical radiative efficiency (Table \ref{tbl-calculated}), suggesting that
the simulation of extreme SSC emission employed therein is not an appropriate representation of the data.  

Models B2 and B3 are also well described by a large influx of material injected into the base of the jet, which may strengthen the effects of particle acceleration at a shock while temporarily weakening the comparative power of the magnetic field, without initiating widespread reconnection.  It would be particularly interesting to study high-energy polarization in blazar flares, especially as several new instruments are in the planning stages. X-ray and $\g$-ray polarimetry can more definitively separate leptonic from hadronic models, elucidate the shock versus magnetic reconnection debate, and provide a new avenue through which to explore the geometry of the acceleration/emission region \citep[e.g.][]{dreyer19}. 
   
\subsection{Comparison with Previous Work}

Besides reporting on the 2013 December Epochs that we model here (and
other epochs that we do not consider), \citet{hayash15} also modeled
these epochs with the {\tt BLAZAR} code \citep{moderski03}.  They used
a double broken power-law ED to model Epoch A, and
a broken power-law distribution to model Epoch B.  Similar to our
modeling, \citet{hayash15} found that EC-dust dominated the $\g$-ray
emission during Epoch A and EC-BLR dominated during Epoch B.  For
Epoch A, they found a larger magnetic field, lower $\G$, and larger
$r_{\rm blob}$.  They modeled Epoch B with two sets of parameters.
Their model parameters for this epoch are generally similar to ours
for our models B2 and B3, although or magnetic field for our model B1
is quite lower compared to their models.  Our models for Epoch B
have similar $r_{\rm blob}$, but larger $\G$ for our models B2 and B3.

\citet{asano15} model the 2013 December flare with a time-dependent model \citep{asano14} that treats particle acceleration quite similar to ours. They model both the SED and the $\g$-ray light curve. In their model, scattering of a UV field (presumably representing a BLR) dominates the $\g$-ray emission and they have $\G=15$ and $r_{\rm blob}\sim 6\times10^{16}\ \cm$, so their emitting region is closer to the BH than in our models (to a degree consistent with our use of a stratified BLR, Paper 2).  Their model successfully reproduces the observed LAT $\gamma$-ray light curve, although the high $\G$ may be a problem for models of jet acceleration by magnetic dissipation.

The 2013 flare was further examined by \citet{paliya16} using
time-dependent lepto-hadronic and two-zone leptonic models.  Those
authors examined a 3 day window that included the flare, and noted a 3
hr flux doubling timescale.  They adopted a smooth broken power-law as
the particle distribution, and found that the isolated flare could
represent $\gamma$-ray emission from a smaller blob, while the rest of
the spectrum was produced in a larger volume. Another possibility is
that proton acceleration could power the enhanced $\g$-ray emission.
The inclusion of significant non-flare data in the analysis of
\citet{paliya16} raises concerns about identifying the physics of the
flare specifically. Hence the shape of the ED and
the nature of the associated particle acceleration mechanism(s) during
the 2013 flare from 3C~279 are still open questions.

\subsection{Summary}

Our model from Paper~2 included a numerical steady-state solution to a
particle transport equation that included first- and second-order Fermi acceleration, and particle escape.  Particles in this single homogeneous blob can lose energy to adiabatic expansion, synchrotron radiation, and inverse-Compton scattering of incident photon fields including SSC, EC/Disk, EC/Dust, and EC/BLR, where the BLR is composed of 26 individual lines radiating at infinitesimally thin concentric shells.  We added to this $\g\g$-absorption due to the accretion disk, dust torus, and stratified BLR consistent with the emission and Compton scattering geometry \citep{finke16}.  We applied our model to the SED of the extremely Compton dominant flare of 3C 279 observed on 2013 December 20 \citet{hayash15}.  The preceding 3 days of quiescent data are similarly analyzed as a baseline for the flare simulations. We derive a simplified version of the Thomson regime approximation (Paper~1), which assists in the interpretation of the ED.  Our primary results are as follows:

\begin{itemize}
\item It is possible to simulate the acceleration in the flare SED
  with reasonable levels of only first- and second-order Fermi processes (with
  the latter process dominating in our models).  Acceleration by
  reconnection is not needed, and the maximum magnetization parameter
  ($\sigma$) found from our model parameters is consistent with this.

\item It is possible for the BLR to be the dominant EC component without significant $\g\g$ absorption
from BLR photons.  

\item The quiescent period displays electron and field powers near equipartition, while the flare is strongly electron dominated. 

\item There is insufficient energy available for the X-rays to have undergone a flare comparable to the $\g$-ray flare simultaneously. 

\item The simplified ED analysis clarifies that
  first- and second-order Fermi acceleration can influence different components
  of the overall ED.

\item Based on our modeling of Epoch B, the $\nu F_\nu$ flux at
  $\approx10$\ MeV is likely $< 10^{-9}\ \erg\ \s^{-1}\ \cm^{-2}$
  (i.e., Model B1 is highly unlikely).  The emission in this energy
  range could be probed by a future $\g$-ray mission such as
  e-Astrogam or AMEGO, which could provide further constraints on
  blazar SEDs.

\end{itemize}

This model has been informative in the analysis of this particularly unusual flare.  We plan to use the same model to analyze other intriguing blazar behaviors.  Additionally the simplified analytic ED can be applied to astrophysical jets more broadly, where both first- and second-order Fermi acceleration are expected to contribute. 

\acknowledgements 

We thank the anonymous referee for insightful comments, which improved the presentation and clarity of the manuscript.  We thank Masaaki Hayashida for furnishing the SED data for our analysis. T.R.L was partially supported by a George Mason University
Dissertation Research Grant and the Zuckerman Institute as a Zuckerman
Postdoctoral Scholar. J.D.F. was supported by NASA under contract
S-15633Y.

\clearpage

\appendix

\section{Derivation of the Thomson Electron Distribution Features}
\label{ap-bump}

In Paper~1, the electron transport equation is solved analytically in the Thomson limit.  That steady state analytic solution (Equation \ref{eq-analytic})
\begin{flalign}
\Ngreen(\g) = \frac{\dot{N}_{\rm inj} e^{b\g_{\rm inj}/2}}{bD_0 \g_{\rm inj}^{2+(a/2)}} \frac{\G[\mu - \kappa +1/2]}{\G[1+ 2\mu]} e^{-b\g/2}\g^{a/2} 
\begin{cases}
{\mathcal M}_{\kappa,\mu}(b\g) {\mathcal W}_{\kappa,\mu}(b\g_{\rm inj}), \g \le \g_{\rm inj}\\
{\mathcal M}_{\kappa,\mu}(b\g_{\rm inj}) {\mathcal W}_{\kappa,\mu}(b\g), \g \ge \g_{\rm inj}
\end{cases} \ ,
\label{eq-anasoln-a}
\end{flalign}
introduces the idea that the ED shape can be interpreted physically if it comes from first-principles.  In this appendix, that solution is simplified by making some mathematical approximations for the phase spaces applicable to the physical regimes of interest for blazars.  The simplified functions provided, may provide insight into a broader range of blazar activity in consistent parameter spaces, as well as any astrophysical jet for which the same assumptions are valid.  

The ED as stated in Equation (\ref{eq-analytic}) for the steady-state solution is branched, with continuity enforced at the injection energy.   The derivation begins with the branch above the injection energy, $\g_{\rm inj}<\gamma$  
\begin{equation}
N(\gamma) = \frac{\dot{N}_{\rm inj}}{D_0} b^{-1} \gamma^{a/2} \g_{\rm inj}^{-(a/2)-2} {\rm exp}\left[ \frac{b\g_{\rm inj}}{2} \right]  {\rm exp}\left[ -\frac{b\gamma}{2} \right]  \frac{\Gamma[1/2 + \mu - \kappa]}{\Gamma[1+2\mu]} {\mathcal M}_{\kappa, \mu} (b \g_{\rm inj}) {\mathcal W}_{\kappa,\mu}(b\gamma) \ .
\label{eq-anasoln-b}
\end{equation}
because the injection energy is almost always smaller than the energies of interest in the rest of the ED, since the injection energy originates from a thermal distribution before arriving in the emitting blob.

The Whittaker functions are replaced with their confluent hypergeometric counterparts 
\begin{align}
N(\gamma) =& \frac{\dot{N}_{\rm inj}}{D_0} b^{-1} \gamma^{a/2} \g_{\rm inj}^{-(a/2)-2} {\rm exp}\left[ \frac{b\g_{\rm inj}}{2} \right]  {\rm exp}\left[ -\frac{b\gamma}{2} \right]  \frac{\Gamma[1/2 + \mu - \kappa]}{\Gamma[1+2\mu]} \\ \nonumber
&\times {\rm exp}\left[ \frac{-b\g_{\rm inj}}{2} \right] b^{(1/2) + \mu} \g_{\rm inj}^{(1/2) + \mu} M[1/2 + \mu - \kappa, 1+2\mu, b\g_{\rm inj}] \\ \nonumber
&\times {\rm exp}\left[ \frac{-b\gamma}{2} \right] b^{(1/2) + \mu} \gamma^{(1/2) + \mu} U[1/2 + \mu - \kappa, 1+2\mu, b\gamma] \ ,
\end{align}
\citep{slater60} and the electron number distribution is simplified,
\begin{equation}
N(\gamma) = \frac{\dot{N}_{\rm inj}}{D_0} b^{a+3} \gamma^{a+2} {\rm exp}[-b\gamma] \frac{\Gamma[1/(b\tau)]}{\Gamma[a+4]}  M[1/(b\tau), a+4, b\g_{\rm inj}]  U[1/(b\tau), a+4, b\gamma] \ ,
\end{equation}
where the Whittaker coefficients are
\begin{flalign}
\kappa=2 - \frac{1}{b\tau} + \frac{a}{2},\quad {\rm and } \quad \mu=\frac{a+3}{2} \ ,
\end{flalign}
as stated after Equation (\ref{eq-analytic}).

The exact expression for the confluent hypergeometric $U$-function is given by
\begin{equation}
U[\hat{a},\hat{b},\hat{x}] = \frac{\pi}{{\rm sin}[\pi \hat{b}]} \left( \frac{M[\hat{a},\hat{b},\hat{x}] }{\Gamma[1+\hat{a}-\hat{b}]\Gamma[\hat{b}]} -  \frac{\hat{x}^{1-\hat{b}} M[\hat{a},\hat{b},\hat{x}] }{\Gamma[\hat{a}]\Gamma[2-\hat{b}]}  \right) \ ,
\label{eq-Usplit}
\end{equation}
\citep{slater60,abramowitz72} which is undefined for nonpositive integer inputs to the $\G$-functions.

In comparing the model with blazar SED data, the combined emission coefficient is small ($b \sim 10^{-6}$), which makes the final argument in the confluent hypergeometric functions small at most Lorentz factors for which there is a meaningful number of electrons ($\g_{\rm max} \lesssim b^{-1}$).  When $b\g \ll 1$ is assumed, the confluent hypergeometric $M$-function can be approximated as
\begin{equation}
M[\hat{a},\hat{b},\hat{x} \rightarrow 0] = 1 \ 
\label{eq-Mappx}
\end{equation}
\citep{slater60,abramowitz72}.  Thus, the electron number distribution can be expressed as
\begin{align}
\label{apeq-8}
N(\gamma) = \frac{\dot{N}_{\rm inj}}{D_0} b^{2\mu} \gamma^{a+2} {\rm exp}[-b\gamma] &\left( \frac{\pi}{{\rm sin}[\pi (a+4)]}  \frac{\Gamma[(b\tau)^{-1}] }{\Gamma^2[a+4]\Gamma[(b\tau)^{-1} -a-3]} \right. \\
&\quad \left. -  \frac{\pi}{{\rm sin}[\pi (a+4)]}  \frac{(b\g)^{-a-3}}{\Gamma[a+4]\Gamma[-a-2]}  \right) \ , \nonumber
\end{align}
with appropriate substitutions, and incomplete factoring which becomes convenient.  It will also become convenient to name the first term on the right hand side of the equation (RHS-1) and the second term on the right hand side of the equation (RHS-2).  

Both terms (RHS-1 and RHS-2) can be simplified, using combinations of the reflection 
and recursion relations for $\G$-functions, 
\begin{align}
\G[z] \G[1-z] = \frac{\pi}{{\rm sin}[\pi z]} \ ,
\label{apeq-9}
\end{align}
and 
\begin{align}
\G[z+1] = z \G[z] \ ,
\label{apeq-10}
\end{align}
respectively \citep{abramowitz72}.

Applying Equation (\ref{apeq-9}) and Equation (\ref{apeq-10}) to the ED in Equation (\ref{apeq-8}), the solution is simplified to
\begin{equation}
N(\gamma) = \frac{\dot{N}_{\rm inj}}{D_0}  \gamma^{-1} {\rm exp}[-b\gamma]  \left( \frac{(b\g)^{a+3}\Gamma[(b\tau)^{-1}]\Gamma[-a-3]}{\Gamma[a+4] \Gamma[(b\tau)^{-1} - a - 3]}  + \frac{1}{3+a}  \right) \ .
\label{apeq-11}
\end{equation}
In each blazar and epoch examined in this work and previous analyses (Papers 1 and 2) the Bohm timescale $\tau \gg b^{-1}$.  It follows that $(b\tau)^{-1} \ll |a+3|$, and adopting this approximation, the $\G$-functions can be simplified as
\begin{equation}
\G[(b\tau)^{-1}-a-3] \approx \G[-a-3] \qquad {\rm and} \qquad \G[(b\tau)^{-1}] \approx b\tau \ ,
\label{apeq-12}
\end{equation}
since $\G^{-1}[z] = z + \mathcal{O}(z^2)$ for $z \ll 1$ \citep{abramowitz72}.  Applying the approximation in Equation (\ref{apeq-12}) to the ED in Equation (\ref{apeq-11}) gives
\begin{equation}
N(\gamma) = \frac{\dot{N}_{\rm inj}}{D_0} {\rm exp}[-b\gamma]  \left( \frac{\tau b^{a+4} \g^{a+2}}{\Gamma[a+4]}  + \frac{\g^{-1}}{3+a}  \right) \ .
\label{eq-splitU}
\end{equation}
This equation was tested for a range of parameter values, and found to agree completely with the analytic solution for parameters consistent with the stated assumptions, namely $\g>\g_{\rm inj}$, $b\g \ll 1$ and $(b\tau)^{-1} \ll 1$. 

\begin{figure}
\vspace{2.2mm} 
\centering
\includegraphics[width=0.5\textwidth]{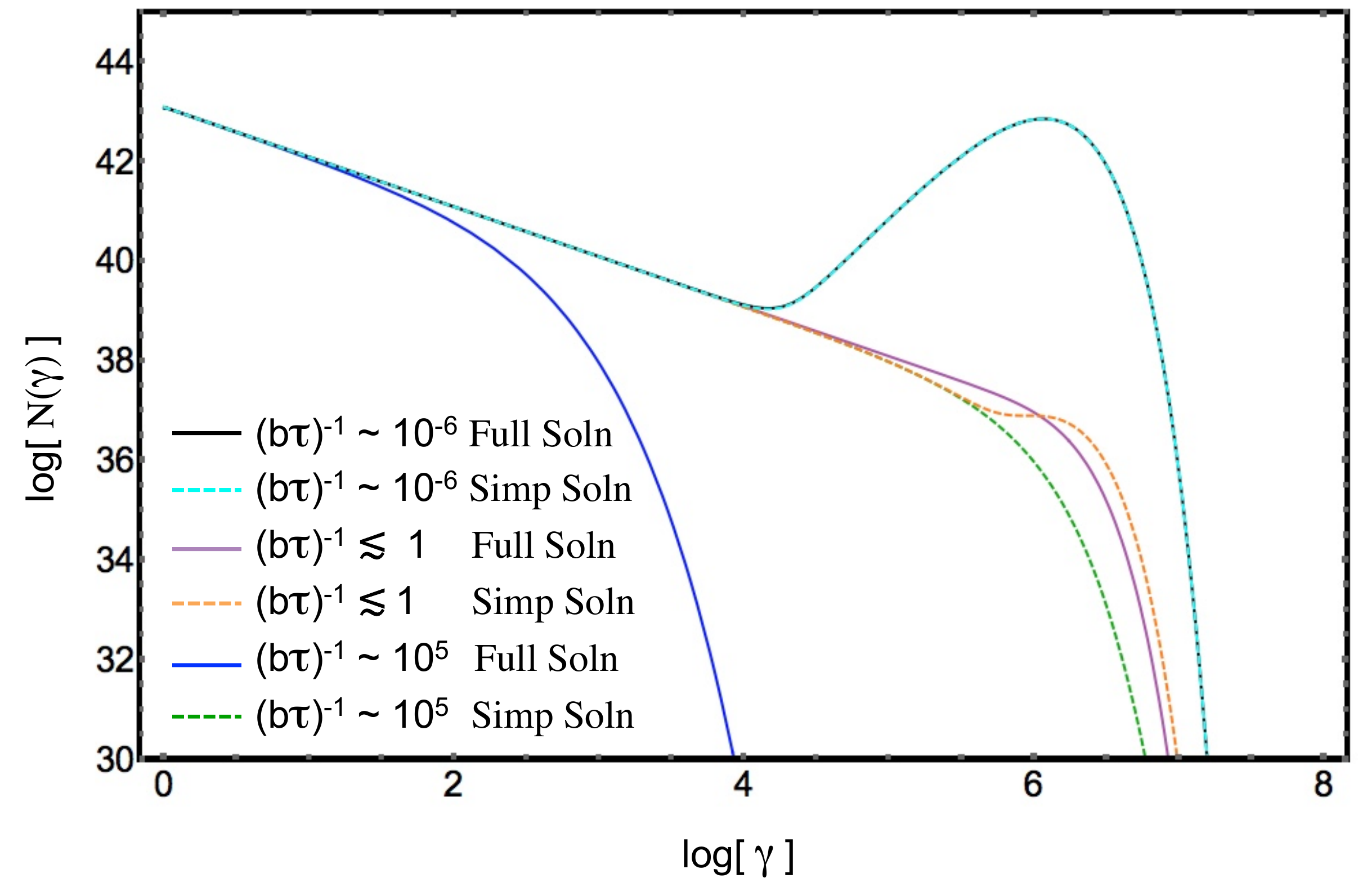}
\caption{The simplified ED (Equation \ref{eq-splitU}) is more sensitive to the limit of $(b \tau)^{-1} \ll 1$ than to $b\g \ll 1$, when compared with the full Thomson ED (Equation \ref{eq-anasoln-b}) in the $\g_{\rm inj} < \g$ regime. }
\label{fig-AppxLimitsTest}
\vspace{2.2mm}
\end{figure}

Figure \ref{fig-AppxLimitsTest} explores the precision of the approximations in the simplified analytic ED (Equation \ref{eq-splitU}) in comparison to the full analytic Thomson ED (Equation \ref{eq-anasoln-b}) for $\g_{\rm inj} < \g$.   The parameters used in this example are similar to those from Model B1 (Table \ref{tbl-freeparams}), however that model formally included the Klein-Nishina cross-section.  Mathematically, we assume that $b\g \ll 1$ in order to use the approximation in Equation (\ref{eq-Mappx}), however at $\g \sim 10^7$, $b\g \sim 25$ for the case where $(b\tau)^{-1} \ll 1$ (black and cyan), and there is no discernible difference between the simplified and Thomson solutions.  Attempts at increasing the value of $b$, generally resulted in both solutions experiencing exponential drops at a comparably smaller Lorentz factor $\g$.  Thus, the approximation is fairly robust.  Conversely, Figure \ref{fig-AppxLimitsTest} demonstrates that the simplified ED is more sensitive to the limit $(b\tau)^{-1} \ll 1$ required for Equation (\ref{apeq-12}).  When the limit is satisfied, the simplified and Thomson solutions are in agreement (black and cyan), and where $(b\tau)^{-1} \gg 1$, the solutions diverge.  In the case of $(b\tau)^{-1} \lesssim 1$, the simplified solution is visibly different, but perhaps sufficient for some applications.  

\begin{figure}
\vspace{2.2mm} 
\centering
\includegraphics[width=0.5\textwidth]{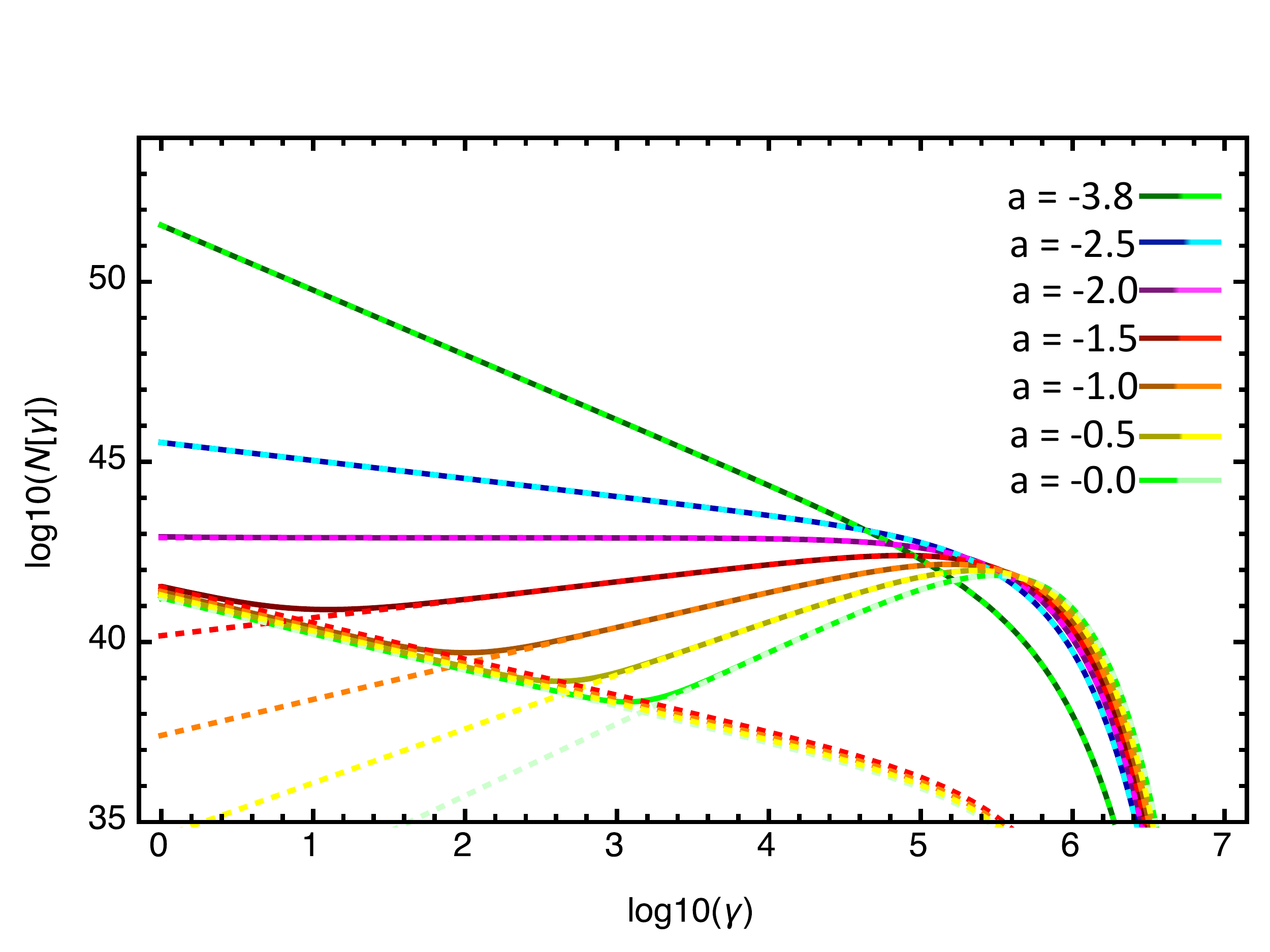}
\caption{Electron distribution for changing $a$ values, where dotted lines are the separate terms in Equation (\ref{eq-splitU}).  The dotted curves for RHS-1 are those which dominate the ED at higher energies, while the doted curves for RHS-2 are shown in the same color scheme.  The dotted curves for RHS-2 are only shown if they have a region of dominance (at lower energies) in the overall ED; otherwise they were suppressed for clarity because they crowded the existing RHS-2 curves.  }
\label{fig-EDsplitU-basic}
\vspace{2.2mm}
\end{figure}

Equation (\ref{eq-splitU}) is shown in Figure \ref{fig-EDsplitU-basic} for a sample set of parameters that illustrate the features well, and wherein only the first-order Fermi/adiabatic parameter $a$ is varied, while all other parameters are fixed.  The high-energy turnover in the ED is controlled by the leading exponential $e^{-b\g}$.  RHS-1 from Equation (\ref{eq-splitU}) can be the dominant term throughout, especially for more negative values of $a$ in this example, but also depending on $b$ and $D_0$ more generally.  For less negative (or increasingly positive) values of $a$, sometimes, RHS-2 becomes the dominant term at low energies, depending on the relative magnitude of each power-law.  RHS-2 is defined by a $\g^{-1}$ power-law in all cases, until the exponential turnover.  RHS-1 is defined by a $\g^{a+2}$ power-law, which is why the lower energy slope of this curve changes with $a$, and always produces a horizontal feature for $a=-2$, regardless of the other parameters.

In this paper, we argue against using substantial injection energies.  Thus, all of our EDs are essentially calculated for $\g \ge \g_{\rm inj}$.  However, the original steady-state analytic solution (Paper 1) did not carry that constraint.  So, for completeness, in the case where $\g \le \g_{\rm inj}$, the ED is given by 
\begin{equation}
N(\gamma) = \frac{\dot{N}_{\rm inj}}{D_0} b^{-1} \gamma^{a/2} \g_{\rm inj}^{-(a/2)-2} {\rm exp}\left[ \frac{b\g_{\rm inj}}{2} \right]  {\rm exp}\left[ -\frac{b\gamma}{2} \right]  \frac{\Gamma[1/2 + \mu - \kappa]}{\Gamma[1+2\mu]} {\mathcal M}_{\kappa, \mu} (b \g) {\mathcal W}_{\kappa,\mu}(b\g_{\rm inj}) \ .
\end{equation}
Following the same arguments presented above, the ED can be simplified to 
\begin{equation}
N(\gamma) = \frac{\dot{N}_{\rm inj}}{D_0} {\rm exp}[-b\gamma]  \left( \frac{\tau b^{a+4} \g^{a+2} }{\Gamma[a+4]}  + \frac{\g_{\rm inj}^{-a-3} \g^{a+2} }{3+a}  \right) \ .
\label{eq-splitU-x0}
\end{equation}
In this approximate ED, RHS-1 is identical to RHS-1 in Equation (\ref{eq-splitU}), and essentially makes no contribution. In Equation (\ref{eq-splitU-x0}), RHS-2 normalizes the ED to the appropriate magnitude for the injection energy, and $\g^{a+2}$ still governs the shape.  

\begin{figure}
\vspace{2.2mm} 
\centering
\includegraphics[width=0.5\textwidth]{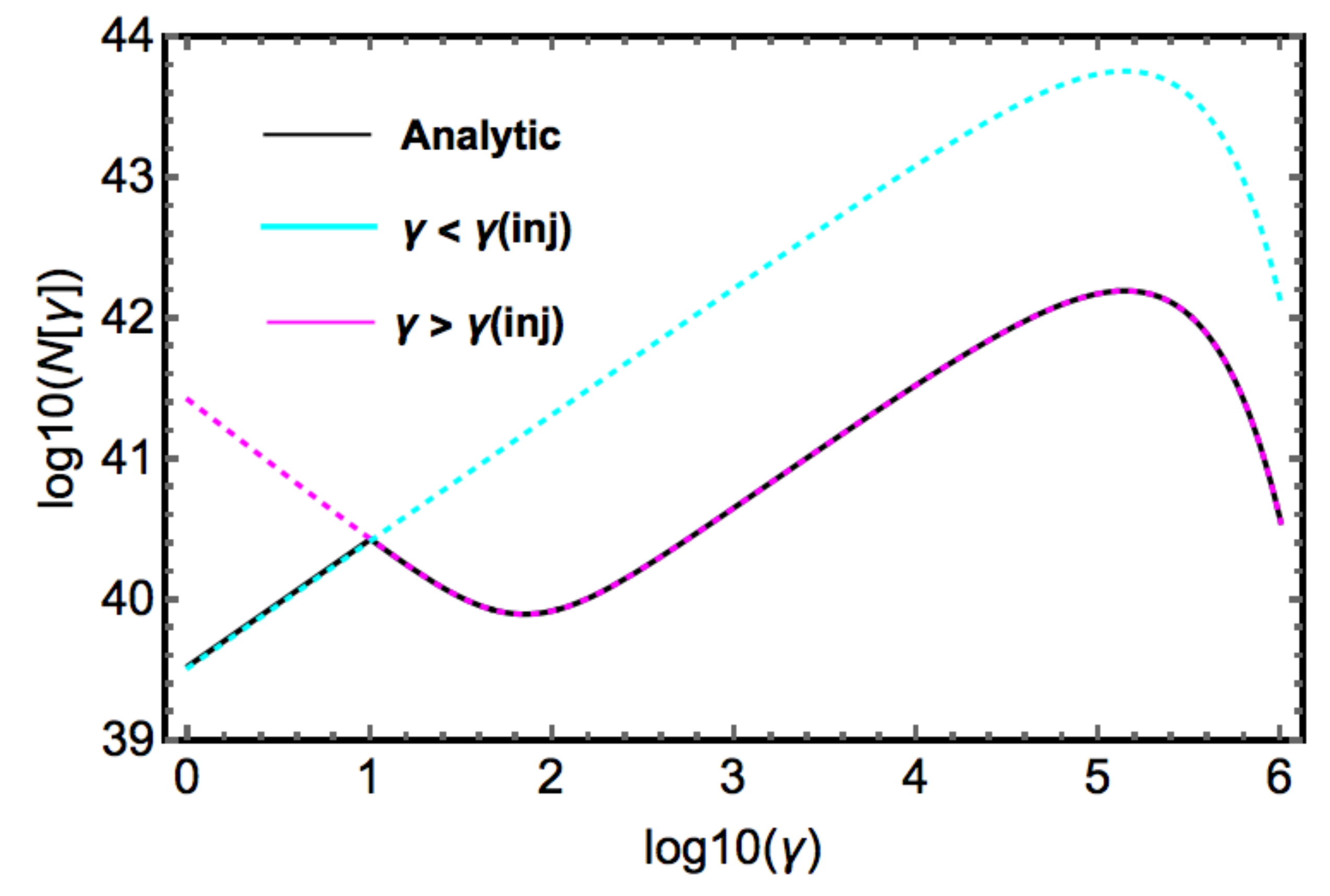}
\caption{Electron distribution for $a=-1$, where we let $\g_{\rm inj} = 10$ in order to observe the injection energy cusp.  The underlying black line is the full analytic ED shown in Equation (\ref{eq-analytic}) of Section \ref{model} (Paper 1).  The magenta dotted line is the solution from Equation (\ref{eq-splitU}) for $\g \ge \g_{\rm inj}$, while the cyan line is the solution in Equation (\ref{eq-splitU-x0}) for $\g \le \g_{\rm inj}$.  }
\label{fig-EDsplitU-basic}
\vspace{2.2mm}
\end{figure}

For physical SED evaluations, the analysis tends to concentrate in the regions of parameter space where RHS-1 in Equation (\ref{eq-splitU}) dominates.  This expression could be helpful to those who assume a power-law ED with an exponential cutoff, since it provides a physical interpretation behind that general shape including first-order Fermi and diffusive acceleration, as well as synchrotron and Thomson emission. However the analysis of Epoch B demonstrates the usefulness of the full simplified solution in Equation (\ref{eq-splitU}) for variety of ED shapes.

To further expound upon Equation (\ref{eq-splitU}), the leading factor $e^{-b\g}$ provides a turnover in the ED at high energies, which is dependent upon the non-thermal cooling coefficients for synchrotron radiation, and Compton scattering.  Note the first term on the right hand side of Equation (\ref{eq-splitU}), where $N(\g) \propto \g^{a+2}$ implies that where this term dominates, the lower energy slope is defined by $(a+2)$, where the first-order Fermi acceleration/adiabatic expansion coefficient $a$ is often negative in blazar spectral analysis (Papers 1 and 2), and will always be flat for $a=-2$.  

There are parameter regimes, especially in the samples provided in Figure \ref{fig-EDsplitU-basic}, where $a > -2$, and it is possible to see the second term in Equation (\ref{eq-splitU}) become dominant in the lower energy domain (but still above the injection energy).  In this scenario, the change occurs where the two terms in Equation (\ref{eq-splitU}) become equivalent, or at the Lorentz factor
\begin{equation}
\g = \g_{\rm dip} = \frac{1}{b} \left( \frac{\G[a+3]}{b\tau} \right)^{1/(a+3)} \ ,
\end{equation}
as is the case for the Epoch B analysis in Section \ref{3c279}.

For $\g_{\rm inj} < \g < \g_{\rm dip}$, if RHS-2 dominates in Equation (\ref{eq-splitU}), then the electron population is dominated by diffusion from the injection energy.  The steady-state solution contains a continuous, monochromatic particle injection, and the transport equation allows the electron population to evolve over time, where the steady-state is an equilibrium snapshot.  The second term (RHS-2) electrons are those which did not have time to get sufficiently caught up in the acceleration processes to lose the signature of their initial injection.  

For $\g_{\rm dip} < \g < 1/b_{\rm tot}$, RHS-1 of Equation (\ref{eq-splitU}) tends to dominate.  In this domain, acceleration and diffusion processes are in perfect balance with one another, and there is no net transport.  However, it is possible for the ``acceleration'' portion to be negative due to adiabatic expansion, hence electrons are ``accelerated'' to lower energies.  When acceleration $a$ is stronger (more positive), $\g_{\rm dip}$ is pushed to higher energies, as the electrons form an increasingly bimodal distribution (see Figure \ref{fig-EDsplitU-basic}).

\section{Acceleration Term Dependences in the Simplified Solution}
\label{ap-deps}

The electron number distribution is defined according to the distribution function (Paper~1), 
\begin{align}
N =4\pi (m_e c)^3 \g^2 f(\g) \ ,
\end{align}
for the steady state case.  Consider the domain of the simplified ED where $\g > \g_{\rm dip} > \g_{\rm inj}$ (Equation \ref{eq-splitU})
\begin{align}
N \propto \g^{a+2} \ .
\end{align}
It follows that the distribution function follows the proportionality 
\begin{align}
f \propto \g^a\ ,
{\rm or } \quad f(\g) = A_*\g^a
\end{align}
where $A_*$ is a place-holding constant.  The particle flux due to stochastic diffusion (a second-order process; Equation 59 of Paper 1)
\begin{equation}
\dot{N}_{\rm sto} = -aD_0A_*\g^{a+3} \ ,
\end{equation}
and the particle flux due to first-order Fermi acceleration (including adiabatic expansion)
\begin{equation}
\dot{N}_{\rm ad+sh} = A_{\rm ad+sh} \g^3 f = A_{\rm ad+sh}A_*\g^{a+3} \ ,
\end{equation}
can be combined into the particle flux due to the sum of first- and second order Fermi acceleration
\begin{align}
\dot{N}_{\rm sum} &= \dot{N}_{\rm sto} + \dot{N}_{\rm ad+sh} = -aD_0A_* \g^{a+3} + A_{\rm ad+sh}A_*\g^{a+3} \equiv 0 \ , 
\end{align}
since $A_{\rm ad+sh} \equiv aD_0$, demonstrating that these processes perfectly balance in all cases for RHS-1 of Equation (\ref{eq-splitU}).

There are blazar data analyses where the first-order Fermi coefficient $a < -2$, and it is possible to see the second term in Equation (\ref{eq-splitU}) become dominant in the lower energy domain (but still above the injection energy).  In this scenario, the change occurs where the two terms become equivalent
\begin{align}
\frac{b^{a+3}\g^{a+2}\G(1/(b\tau))}{\G(a+4)} = \frac{1}{\g(a+3)} \ , 
\end{align}
which can be solved for the Lorentz factor where the two terms cross, 
\begin{equation}
\g = \g_{\rm dip} = \left[ \frac{1}{(a+3)b^{a+3}} \frac{\G(a+4)}{\G(1/(b\tau))} \right]^{1/(a+3)} \ .
\end{equation}

For $\g_{\rm inj} < \g < \g_{\rm dip}$, if the second term dominates, then the electron population is dominated by diffusion from the injection energy.  The continuous particle injection to the transport equation allows the electron population to evolve over time, where the steady-state is an equilibrium snapshot.  The second term represents electrons that did not have time to get sufficiently caught up in the first-order Fermi acceleration processes to lose the signature of their initial injection.  

The ED where RHS-2 of Equation (\ref{eq-splitU}) dominates is described by the proportion, 
\begin{equation}
N(\g) \propto \frac{1}{\g} \ ,
\end{equation}
meaning that the distribution function is given by (Paper~1)
\begin{equation}
f(\g) \propto \frac{N}{\g^2} \qquad {\rm or } \qquad f(\g) = \frac{A_*}{\g^3} \ .
\end{equation}
The particle flux due to second-order Fermi diffusion is given by (Equation 59 of Paper~1)
\begin{align}
\dot{N}_{\rm sto} &= -D_0\g^{4} \frac{\partial f}{\partial \g} = 3D_0 A_* \ .
\end{align}
and the particle flux due to first-order Fermi acceleration (including adiabatic expansion) is
\begin{align}
\dot{N}_{\rm ad+sh} &= A_{\rm ad+sh} \g^3 f = A_{\rm ad+sh} A_* \ .
\end{align}
The rates of first- and second order Fermi acceleration can be combined into the total rate
\begin{align}
\dot{N}_{\rm sum} &= \dot{N}_{\rm sto} + \dot{N}_{\rm ad+sh} = A_* D_0 (3 + a) \ , 
\end{align}
where the first-order Fermi acceleration and second-order Fermi diffusion processes are not perfectly balanced for all parameters, but can balance where $a=-3$.  As it happens, this is similar to first-order Fermi acceleration/adiabatic expansion parameters we find when comparing to data.  Specifically, $a \sim -3.8 \pm 0.3$ Paper~2, indicating that both parameters are necessary.

For $\g_{\rm dip} < \g < 1/b$, the first term tends to dominate.  In this regime, first-order Fermi (with adiabatic expansion) and second-order Fermi acceleration processes are in perfect balance with one another, and there is no net transport.  However, it is possible for the first-order ``acceleration'' portion to be negative due to adiabatic expansion, hence electrons are ``accelerated'' to lower energies.  When first-order Fermi acceleration $A_{\rm ad+sh}$ is stronger (more positive), $\g_{\rm dip}$ is pushed to higher energies, as the electrons form an increasingly bimodal distribution.

It is important to note that while one term may be negligible for specific parameter ranges where another is clearly dominant, no term alone represents a complete, independent solution to the electron transport equation.  Therefore, interpretations of the ED, must be mindful that the solution terms are not completely separable to maintain a self-consistent picture in the original sense of the transport equation.  Recall that we have made a number of assumptions about the ranges of several parameters in order to provide these simplified expressions and their corresponding interpretations.  We anticipate that the simplified function forms will be useful, but we encourage some caution in their application.

\clearpage


\bibliographystyle{apj}
\bibliography{msNote,EBL_ref,references,mypapers_ref,blazar_ref,sequence_ref,SSC_ref,LAT_ref,3c454.3_ref}

\end{document}